%% file: main.tex
\documentclass[journal]{IEEEtran}
\usepackage{cite}

\ifCLASSINFOpdf
\usepackage[pdftex]{graphicx}
\else
\fi

\usepackage{amsmath}
\usepackage{amsfonts}

\usepackage{fixltx2e}

\usepackage{url}

\usepackage{siunitx}  
\usepackage{acronym}  

\acrodef{2d}[2D]{two-dimensional}
\acrodef{3d}[3D]{three-dimensional}
\acrodef{5g}[5G]{5th-generation}
\acrodef{adc}[ADC]{analog to digital converter}
\acrodef{agc}[AGC]{automatic gain control}
\acrodef{aoa}[AoA]{angle-of-arrival}
\acrodef{aod}[AoD]{angle-of-departure}
\acrodef{awgn}[AWGN]{additive white Gaussian noise}
\acrodef{bs}[BS]{base station}
\acrodef{cdf}[CDF]{cummulative distribution function}
\acrodef{cir}[CIR]{channel impulse response}
\acrodef{crlb}[CRLB]{Cram\'{e}r–Rao lower bound}
\acrodef{ctf}[CTF]{channel transfer function}
\acrodef{dmc}[DMC]{dense multipath component}
\acrodef{eadf}[EADF]{effective aperture distribution function}
\acrodef{fim}[FIM]{Fisher information matrix}
\acrodef{fpga}[FPGA]{field-programmable gate array}
\acrodef{gpio}[GPIO]{general-purpose input/output}
\acrodef{gnss}[GNSS]{global navigation satellite system}
\acrodef{gps}[GPS]{Global Positioning System}
\acrodef{imu}[IMU]{inertial measurement unit}
\acrodef{isac}[ISAC]{integrated sensing and communication}
\acrodef{lhf}[LHF]{likelihood function}
\acrodef{lidar}[lidar]{light detection and ranging}
\acrodef{lo}[LO]{local oscillator}
\acrodef{los}[LoS]{line-of-sight}
\acrodef{mimo}[MIMO]{multiple-input multiple-output}
\acrodef{mmse}[MMSE]{minimum mean-square error}
\acrodef{mmwave}[mmWave]{millimeter-wave}
\acrodef{mpc}[MPC]{multipath component}
\acrodef{mrt}[MRT]{maximum ratio transmission}
\acrodef{mse}[MSE]{mean-square error}
\acrodef{nlos}[NLoS]{non-\ac{los}}
\acrodef{ofdm}[OFDM]{orthogonal frequency-division multiplexing}
\acrodef{ota}[OTA]{over-the-air}
\acrodef{papr}[PAPR]{peak-to-average-power ratio}
\acrodef{pcrlb}[PCRLB]{posterior \ac{crlb}}
\acrodef{pdf}[PDF]{probability density function}
\acrodef{pll}[PLL]{phase-locked loop}
\acrodef{pps}[1PPS]{1 pulse per second}
\acrodef{rf}[RF]{radio frequency}
\acrodef{rmse}[RMSE]{root mean square error} 
\acrodef{sdr}[SDR]{software-defined radio}
\acrodef{simo}[SIMO]{single-input-multiple-output}
\acrodef{slam}[SLAM]{simultaneous localization and mapping}
\acrodef{snr}[SNR]{signal-to-noise ratio}
\acrodef{tdma}[TDMA]{time-division multiple access}
\acrodef{ue}[UE]{mobile user}
\acrodef{usrp}[USRP]{Universal Software Radio Peripheral}
\acrodef{wssus}[WSSUS]{wide-sense stationary and uncorrelated scatterers}

\usepackage{booktabs}

\usepackage{pgfplots}
\pgfplotsset{compat=newest}
\usetikzlibrary{plotmarks}
\usetikzlibrary{arrows.meta}
\usepgfplotslibrary{patchplots}
\usetikzlibrary{patterns}
\usepackage{grffile}
\usepackage{amsmath}
\pgfplotsset{plot coordinates/math parser=false}
\newlength\figureheight
\newlength\figurewidt

\usepackage{pdfpages}
\usepackage{xcolor}

\begin{document}
\raggedbottom
%
\title{A Wideband Distributed Massive MIMO Channel Sounder for Communication and Sensing}
%
%
%
\author{Michiel~Sandra, \textit{Student Member, IEEE},
        Christian~Nelson, \textit{Student Member, IEEE},
        Xuhong~Li, \textit{Member, IEEE},\\
        Xuesong~Cai, \textit{Senior Member, IEEE},
        Fredrik~Tufvesson, \textit{Fellow, IEEE},
        and~Anders~J~Johansson, \textit{Member, IEEE}
\thanks{All authors are with the Department
of Electrical and Information Technology, Lund University, Lund, Sweden. (e-mail: \{michiel.sandra, christian.nelson, xuhong.li, xuesong.cai, fredrik.tufvesson, anders$\_$j.johansson\}@eit.lth.se).

This work was supported in part by the Royal Physiographic Society of Lund, in part by the Horizon Europe Framework Programme under the Marie Sk{\l}odowska-Curie grant agreement No. 101059091, in part by the Swedish Research Council (Grant No. 2022-04691), and in part by the Strategic Research Area Excellence Center at Link\"oping--Lund in Information Technology (ELLIIT).\\
Submitted to IEEE Transactions on Antennas and Propagation
}
}

\maketitle

\begin{abstract}
Channel sounding is a vital step in understanding wireless channels for the design and deployment of wireless communication systems.
In this paper, we present the design and implementation of a coherent distributed massive MIMO channel sounder operating at 5-6\,GHz 
with a bandwidth of 400\,MHz based on the NI~USRP~X410. Through the integration of transceiver chains and
RF switches, the design facilitates the use of a larger number of antennas without significant compromise in dynamic capability. Our current implementation is capable of measuring thousands of antenna combinations within tens of milliseconds. Every radio frequency switch is seamlessly integrated with a 16-element antenna array, making the antennas more practical to be transported and flexibly distributed.
In addition, the channel sounder features real-time processing to reduce the data stream to the host computer and increase the signal-to-noise ratio. 
The design and implementation are verified through two measurements in an indoor laboratory environment. The first measurement entails a single-antenna robot as transmitter and 128 distributed receiving antennas. The second measurement demonstrates a passive sensing scenario with a walking person. We evaluate the results of both measurements using the super-resolution algorithm SAGE. The results demonstrate the great potential of the presented sounding system for providing high-quality radio channel measurements, contributing to high-resolution channel estimation, characterization, and active and passive sensing in realistic and dynamic scenarios.


\end{abstract}

\begin{IEEEkeywords}
channel sounding, distributed MIMO, massive MIMO, multistatic radar, 6G, ISAC
\end{IEEEkeywords}

%
\IEEEpeerreviewmaketitle

\section{Introduction}


\IEEEPARstart{C}{oherent} transmission and reception from a massive number of distributed antennas is the foundation of emerging technologies such as distributed massive \ac{mimo} and cell-free massive \ac{mimo} \cite{Ngo2017, Ozlem2021, Ngo2018, Bjoernson2020, Bjoernson2019, Hu2018, Interdonato2019, VanderPerre2019}. These technologies hold the potential to enhance capacity, reliability, localization accuracy and sensing capability.
%
Expanding the number of antennas raises the upper limit on the spatial degrees of freedom of the channel, which could lead to a significant increase in capacity and to higher diversity orders. 
Compared to the conventional \ac{mimo} with antennas in a compact area, separating or distributing the antennas makes the system less prone to large-scale fading effects.
The distributed nature of the antennas also opens up more opportunities for utilizing the channel information for \ac{isac} applications, as the number of physical anchor points increases. 

In order to propose practical solutions and evaluate the system's performance, it is essential to attain a thorough understanding of the distributed massive \ac{mimo} channels. 
Consequently, there is a need to design a channel sounder that can capture the real-world propagation characteristics. 
The channel sounder should have a large number of antennas that can be flexibly distributed, 
a large bandwidth to obtain high delay resolution, and a high snapshot rate to characterize dynamic scenarios.
Moreover, the propagation channels among different \ac{bs} antennas, besides the traditional \ac{bs}-user links, become increasingly important to be understood.
For example, in distributed \ac{mimo} systems, channel state information between \ac{bs} antennas can facilitate calibration procedures that enable coherent transmission and reception \cite{Larsson2023}. Similarly, in \ac{isac}, such channel information can be utilized to detect changes in the environment. Therefore, the channel sounder must be capable of measuring the channels between all combinations of antennas in the system, also known as multi-node or multi-link channel sounding \cite{Nelson2023, Zelenbaba2020, Wassie2019}. Lastly, it is imperative for a channel sounder cost-effective in its implementation and easy to deploy in various scenarios.


In the literature, there are three main ways to increase the number of antennas of a channel sounder: \textit{i)} increasing the number of transceiver chains, \textit{ii)} employing \ac{rf} switches or \textit{iii)} through the virtual array principle.
Several channel sounders have been presented in \cite{Stanko2022, Guevara2021, Laly2015, Euchner2021, Malkowsky2017, Loeschenbrand2019, Loeschenbrand2022} that feature one receiver chain per antenna. The primary limitations of this approach include the increase in cost with the number of antennas, the potential complexity arising from the substantial data generated by the RF chains, and the risk of the sounder becoming cumbersome and unwieldy.
In addition, it can be challenging to perform phase calibration of the transceiver chains, which is needed for \ac{aod} and \ac{aoa} estimation. Nevertheless, a solution was presented in \cite{Loeschenbrand2019} utilizing a custom-made calibration unit.
Channel sounders employing the switched array principle, e.g. \cite{Nelson2024,Flordelis2013, Laly2020}, can address these drawbacks. However, they must compromise on dynamic capability. 
Furthermore, the integration of parallel \ac{rf} channels with \ac{rf} switches, as shown in \cite{Laly2020}, presents a viable trade-off solution.
Channel sounders that employ the virtual array principle \cite{Alatossava2008} -- a convenient and cost-effective method to measure the channel with an immense number of antennas lack dynamic capability.

The bandwidth of a channel sounder is limited by the underlying \ac{rf} equipment. 
Channel sounders based on a vector network analyzer (VNA) \cite{Choi2020} can have a bandwidth on the order of several\,GHz. However, VNAs measure the channel much slower than other solutions and hence limit the dynamic capability. 
Therefore, many channel sounders are based on \acp{sdr} \cite{Stanko2022, Malkowsky2017, Loeschenbrand2019, Guevara2021, Nelson2024, Wassie2019} such as the USRP series from National Instruments, and custom designs \cite{Laly2015, Laly2020, Simon2023, Flordelis2013}. While customs design are typically with high-cost \cite{Laly2015,Laly2020}, software-radios can offer a complete radio system without a high development cost. Naturally, the bandwidth of the channel sounder is limited by the specifications of the state-of-the-art commercially available \acp{sdr}. One also has to consider the data rate between an \ac{sdr} and the host computer, which can easily become several gigabytes per second in (ultra-)wideband setups. When bandwidth is increased, real-time processing on the SDR might be needed to reduce the data rate between the \ac{sdr} and the host computer.

To the best our knowlegde, a channel sounder combing the multi-link capability with a large number of antennas and a large bandwidth has not previously been presented in the literature. Nevertheless, such a channel sounder opens up opportunities to enhance distributed \ac{mimo} channel models, phase calibration techniques, localization algorithms and \ac{isac} methods.
To fill this gap, we present the design, implementation and verification of a scalable multi-link distributed massive \ac{mimo} channel sounder operating in the 5$-$6\,GHz range with an instantaneous bandwidth up to 400\,MHz based on the NI USRP X410. The sounder is capable of measuring the channels between thousands of antenna combinations in tens of milliseconds. 
The main contributions of this work are:
\begin{itemize}
     \item Through the integration of both parallel \ac{rf} chains and RF switches, we facilitate the utilization of a larger number of antennas without significant compromise in dynamic capability. Every SP16T RF switch is seamlessly integrated with a 16-element antenna array within a single panel. This integration streamlines the calibration process in an anechoic chamber, guaranteeing phase coherency among the co-located antennas.
    \item The high data rate problem, due to the 400\,MHz bandwidth, is addressed with real-time processing on the \ac{fpga}, selecting and averaging sounding frames. As a result, we realize high processing gains without increasing the data rate.
    \item The design and implementation of the channel sounder are verified by two sample measurements in an indoor laboratory environment. The two measurements are tailored for a communication application and a sensing application, respectively, where the potential of distributed \ac{mimo} is also well demonstrated. 
\end{itemize}
An early version of the channel sounder, which could only measure a simple single-input single-output link, has been presented in \cite{Sandra2022b}. 
It is worth noting that our design and implementation is open-source \cite{openucs}, based on open-source tool chains, built with off-the-shelf electronic components, allowing other researchers to build a similar channel sounder.


The rest of the paper is organized as follows. Sect.\,\ref{sect2} discusses the channel sounder design and implementation. Sect.\,\ref{sect3} elaborates on the post-processing and calibration aspects. In Sect.\,\ref{sect4}, the two sample measurements and the results that demonstrate the capability of the sounder and the potential of distributed MIMO in communication and sensing are included. Finally, conclusive remarks are summarized in Sect.\,\ref{sect5}.

\emph{Notations}: Throughout this paper, italic letters are used to represent scalars. Bold lowercase letters are for vectors and bold uppercase letters for matrices or tensors. The vector denoted as $\mathbf{1}$ consists of elements that are all equal to one and has proper length. $(\cdot)^{\rm T}$ indicates matrix transposition, $\mathrm{j} = \rm \sqrt{-1}$, $\mathrm{Tr}(\cdot)$ denotes the trace operator, and $\otimes$ represents Kronecker product. $|\cdot|$ and $\angle$ denote the absolute value and the argument of a complex number, respectively. In addition, $\mathrm{max}\{\cdot\}$ and $\mathrm{sum}\{\cdot\}$ denote the maximum value and the sum of all the elements in a vector, respectively.

\section{Design and Implementation} \label{sect2}
\subsection{High-level Design}
\begin{figure}
    \centering
    \includegraphics[width=\columnwidth]{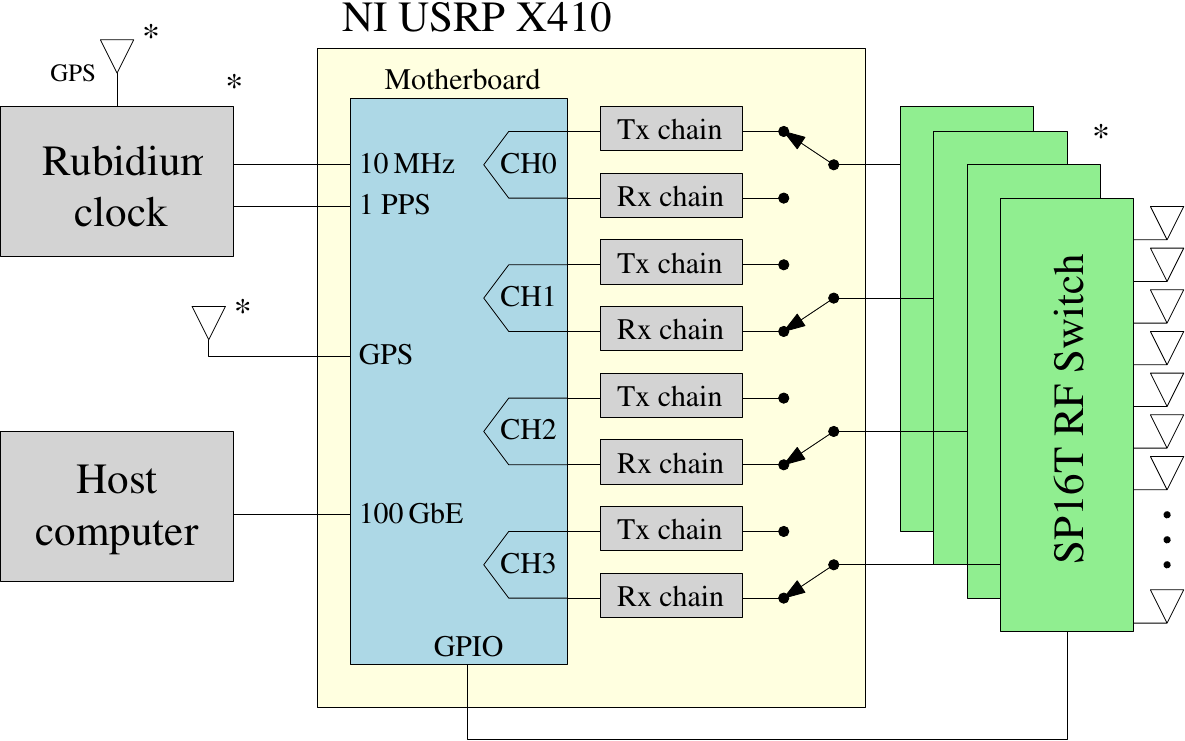}
    \caption{Simplified block diagram of the channel sounder. Components indicated with * are optional.}
    \label{fig:top_block_diagram}
\end{figure}

Fig.~\ref{fig:top_block_diagram} presents a basic block diagram of the channel sounder. Core components include one or more NI USRP X410 and a host computer. Optionally, a Rubidium clock, \ac{gps} antenna, and an RF switch can be integrated into the setup.
The NI USRP X410 is a four-channel full-duplex \ac{sdr}. It has a tunable center frequency ranging from 1 MHz to 7.2 GHz and supports instantaneous bandwidths up to 400 MHz. This device uses the Xilinx Ultrascale+ RFSoC ZU28DR \ac{fpga}, for which we developed custom code.
We implemented channel sounding functions using a custom signal processing block within the RF network on chip (RFNoC) open source framework that manages the streaming and processing of RF data on the \ac{fpga}. It also allows for the design of custom applications with RFNoC blocks. Our processing block can selectively average received samples, optimizing the data stream to the host computer and improving the \ac{snr}.



The Rx and Tx chains are connected to the same port through an internal \ac{rf} switch which is controlled in real time by our custom block. This means that the respective port can transmit or receive, but not both at the same time. 
The following principle applies: when one transmitting, all the other transceiver chains in the channel sounder are receiving. Then, each port in the channel sounder transmits after one another in a pre-configured order.

The USRP is connected to a host computer via an Ethernet interface. Depending on the data rate requirements between the host and the USRP, the 100 GbE interface or the 1 GbE interface can be utilized. Note that multiple USRPs can be connected to one host computer. Apart from the appropriate Ethernet interface, there is no strict hardware requirement for the host computer. However, the host computer has to be fast enough to handle and store the stream of data from the USRP, which can vary from about 1 MB/s to hundreds of MB/s, depending on the configuration.

Rubidium clocks provide a stable time and frequency synchronization between two or more USRPs without the need for any cabling between them during the measurement, allowing for mobile measurements. Note however that the Rubidium clocks have to be synchronized via a cable before the actual measurements. Sufficient synchronization can also be achieved with the \ac{gps} disciplined oscillator built into the USRP and an external \ac{gps} antenna if the measurement scenario allows for stable \ac{gps} reception.

The external \ac{rf} switches allow for a cost-effective extension of the number of channels that can be measured, enhancing the spatial resolution of the propagation measurements. Their states are real-time and independently controlled by our custom RFNoC block through the \ac{gpio} ports of the USRP. 
We have implemented the following \ac{tdma} scheme to control the state of the switches. 
There are in total $N_\mathrm{RF}$ \ac{rf} chains in the system, with
$\mathbf{n_\mathrm{T}} \in \mathbb{N}^{N_\mathrm{RF} \times 1}$ containing the number of (switched) antennas connected to the \ac{rf} chains in the transmission mode, and
$\mathbf{n_\mathrm{R}} \in \mathbb{N}^{N_\mathrm{RF} \times 1}$ containing the number of switched channels for different \ac{rf} chains in the receiving mode. 
Only one antenna is transmitting at a time.
While one antenna is transmitting, the other antennas, except those connected to the same USRP port, are receiving and the RF switches are being switched until all channels have been measured.
After that, the next antenna transmits, etc.
The total number of channels that can be measured is 
\begin{equation}
    \textbf{1}^\mathrm{T} (\textbf{n}_\mathrm{T} \otimes \textbf{n}_\mathrm{R}) - \mathrm{Tr}(\textbf{n}_\mathrm{T} \textbf{n}^\mathrm{T}_\mathrm{R} )
\end{equation}
including reciprocal channels. The total \textit{unique} number of channels is 
\begin{equation}
    \frac{1}{2}\,\textbf{1}^\mathrm{T} (\textbf{n}_\mathrm{T} \otimes \textbf{n}_\mathrm{R}) - \frac{1}{2}\,\mathrm{Tr}(\textbf{n}_\mathrm{T} \textbf{n}^\mathrm{T}_\mathrm{R} )
. \label{eq:nantcomb}
\end{equation}
Because of the simultaneous reception on different \ac{rf} chains, the amount of time slots needed to measure all the antenna combinations is less than the amount of combinations. In our implementation, the total number of time slots it takes to measure the configured channels is 
$
     \max\{\mathbf{n_\mathrm{R}}\}\,\mathrm{sum} \{\textbf{n}_\mathrm{T}\} .
$
At the time of writing, our implementation of the channel sounder contains three USRP X410, three host computers, three Rubidium clocks, and eight Mini-Circuits USB-1SP16T-83H
SP16T RF switches. This makes it possible to measure to measure 7686 unique channels.


\subsection{Elementary Sounding Frame} 
The principle of channel sounding is as follows: a known waveform is transmitted at the desired carrier frequency. On the receiver side, the received waveform is processed to obtain the channel transfer function and impulse response. This could be done for all antenna combinations to recover the directional properties of the channel.
For the design of the channel sounding waveform, it is important that the waveform has a large bandwidth, a uniform power-spectral density, and high energy \cite{book_molisch}. A large bandwidth allows us to achieve a high delay resolution, while the uniform power-spectral density allows us to estimate the channel at across the whole bandwidth with the same quality. 
Higher energy means a higher \ac{snr}. Since the peak power of the transmit power amplifier is limited, high energy implies a low \ac{papr} of the sounding waveform should be used.
It is convenient to resort to the frequency domain to design the sounding waveform as a multi-tone signal. Such a waveform, $\textbf{s} \in \mathbb{C}^{L \times 1}$ with a length of $L$ samples, can in discrete complex baseband be expressed as
\begin{equation}
    \textbf{s}(n) = \frac{1}{L} \sum_{k=0}^{L-1} \textbf{x}(k) \exp\left(\mathrm{j}2\pi k n / L\right),
\end{equation}
where $\textbf{x}\in \mathbb{C}^{L\times1}$ contains the complex amplitudes for all the tones. The entries of $\textbf{x}$ can now be chosen so that $|\textbf{x}(k)| = 1$ for the bandwidth of interest and $\angle\,\textbf{x}(k)$ optimizes the \ac{papr}.
For our channel sounder, we choose a Zadoff-Chu sequence 
for $\textbf{x}(k)$ since $|\textbf{x}(k)| = 1$ and its (inverse) Fourier transform, if the sequence length is a prime number, is also a Zadoff-Chu sequence, which has good time-continuous \ac{papr} and autocorrelation properties \cite{Chu1972, Cai2023}. 

\begin{figure} 
    \centering
    \includegraphics[width=\columnwidth]{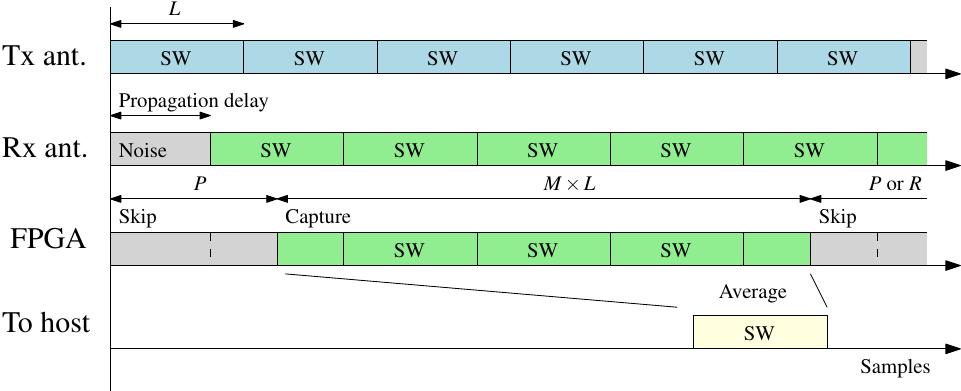}
    \caption{RFNoC processing block functionality for one channel snapshot. SW stands for sounding waveform.}
    \label{fig:sounder}
\end{figure}

During a single-channel snapshot, the respective transmitter sends multiple repetitions of the
same sounding waveform $\textbf{s}$ with a length of $L$ samples, as illustrated in Fig.~\ref{fig:sounder}.
Note that to avoid aliasing, $L$ should be longer than
\begin{equation}
    L \geq \frac{\Delta\tau_\mathrm{max}}{T_\mathrm{s}},
\end{equation}
where $\Delta\tau_\mathrm{max}$ is the maximum delay difference between the first and last received multipath component, and $T_\mathrm{s}$ is the sample period.
These signals then propagate through the radio channel and arrive at
the receiver with an unknown propagation delay. Therefore, samples have to be discarded according to this propagation delay. Otherwise, a couple of samples will be averaged with noise, which reduces the signal-to-noise ratio. In addition, another $\frac{\Delta\tau_\mathrm{max}}{T_\mathrm{s}}$ samples have to be discarded, so the received sounding signals are the result of a circular convolution between the channel impulse response and the original sounding signal. In general, this is not a strict requirement, yet quite convenient for post-processing.
At the beginning of each elementary sounding frame, the RF switches are also configured for the respective antenna; this means that we also have to take into account the switching time of the RF switch.
 We define $P$ as the total number of samples that are discarded and its value should be higher than
\begin{equation}
    P \ge \frac{\tau_\mathrm{0,max}+\Delta\tau_\mathrm{max}+T_\mathrm{sw}}{T_\mathrm{s}}, 
\end{equation}
where $\tau_\mathrm{0,max}$ the maximum delay of the first multipath component, and $T_\mathrm{sw}$ is the maximum time it takes for RF switch to switch channel. The switching time also takes into account the delay of the cable from the \ac{gpio} output of the USRP to the RF switch.
If RF switches are only used for receiving, then the requirement for $P$ could be reduced,
\begin{equation}
    P \ge \mathrm{max}\left(\frac{\tau_\mathrm{0,max}+\Delta\tau_\mathrm{max}}{T_\mathrm{s}}, \frac{T_\mathrm{sw}}{T_\mathrm{s}}\right).
\end{equation}
In practice, $P$ can be set higher than required to account for possible time synchronization errors between receiver and transmitter.
After disregarding $P$ samples, $M$ waveforms of length $L$ are captured and averaged, as illustrated in Fig.~\ref{fig:sounder}.
Averaging is done by averaging each $i^\mathrm{th}$ sample of each received sounding signal. Let us represent the $m^\mathrm{th}$ received waveform as $\mathbf{y}_m \in \mathbb{C}^{L \times 1}$, where $m \in [1,M]$. Then the processed received waveform $\mathbf{y} \in \mathbb{C}^{L\times 1}$ becomes 
\begin{equation}
    {\textbf{y}}(i) = \frac{1}{2^K} \sum^{M}_{m=1} {\textbf{y}_m}(i),
\end{equation}
where $i$ indicates the $i^\mathrm{th}$ element of the vectors, and $K \in \mathbb{N}$ is the parameter for the division factor in the averager. This division is a power of two by design because this simplifies the logic design.
The processed signal is then sent to the host for channel estimation and storage. In the meantime, the RFNoC blocks skips $P$ or $R$ samples until it starts to capture the signal of the next channel. $P$ samples are skipped if the next received signal comes directly after the previous signal. $R$ can be used to adjust the interval $T_\mathrm{rep}$ at which the entire channel is measured. $R$ can be calculated as:
\begin{equation}
    R = \left\lceil\frac{T_\mathrm{rep}}{T_\mathrm{s}}\right\rceil - \max\{\mathbf{n_\mathrm{R}}\}\,\mathrm{sum} \{\textbf{n}_\mathrm{T}\} (P + M L) + P.
\end{equation}
Typical values for the parameters introduced above are presented in Table~\ref{tab:param}.

\subsection{Implementation}
The functionality that was introduced in the previous subsection has been implemented by means of custom logic design on the \ac{fpga} of the X410 and an C++ application on the host computers. These were developed within the open-source RFNoC framework provided by Ettus Research. This framework allows for time-efficient development through so-called RFNoC blocks, similar to the GNU Radio structure. The framework takes care of synchronization, clock generation, data streaming, and analog front-end configuration. By default, each block has an input and output port for both data and control packets. The interface is based on the AXI-Stream protocol. Detailed discussion about the RFNoC framework is beyond the scope of this manuscript and can be found in the USRP Hardware Driver and USRP Manual \cite{usrpmanual}. The complete \ac{fpga} implementation, including all the RFNoC blocks, is called an image. Our image is based on the ``X4\_400" image flavor (UHD v4.2.0.1) and includes two radio blocks, one replay block, and our custom channel sounding block. The radio block is the source and destination block for the RF data and is connected to both the replay block and the channel sounding block. The replay block is an existing block from the RFNoC library and is used to transmit the sounding waveform. With the replay block, we can download our sounding waveform to the DRAM of the X410 and then start transmitting it at the appropriate time. 
When the onboard DRAM is used to transmit the sounding signal, there is no transfer of sample data between the host computer and the USRP. 
Hence, the 1\,GbE Ethernet interface could be used to connect the USRP and the host computer if the USRP would only be used as transmitter or if the receiver data rate is low.

\begin{figure}
    \centering
    \includegraphics[width=0.9\columnwidth]{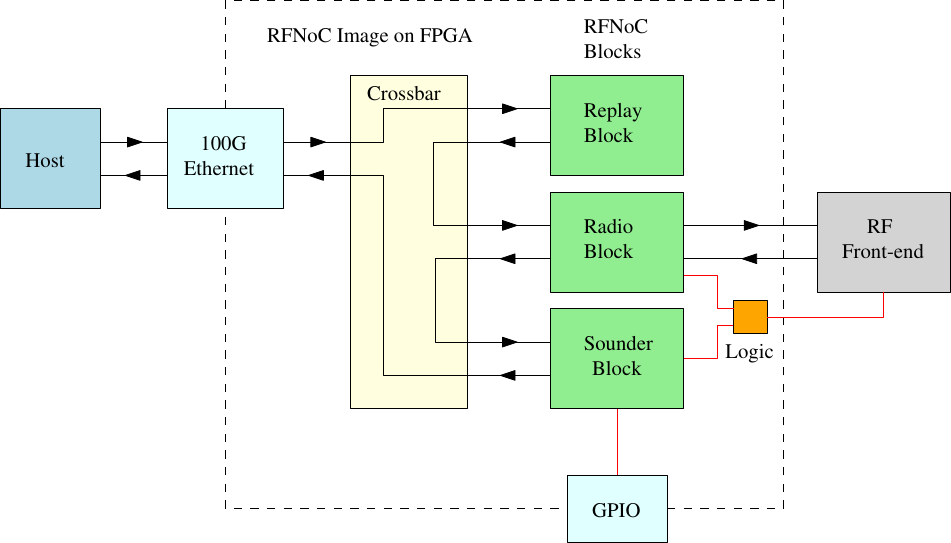}
    \caption{RFNoC image on the FPGA of the USRP X410.}
    \label{fig:arch}
\end{figure}

\begin{figure}[h]
    \centering
    \includegraphics[width=0.6\columnwidth]{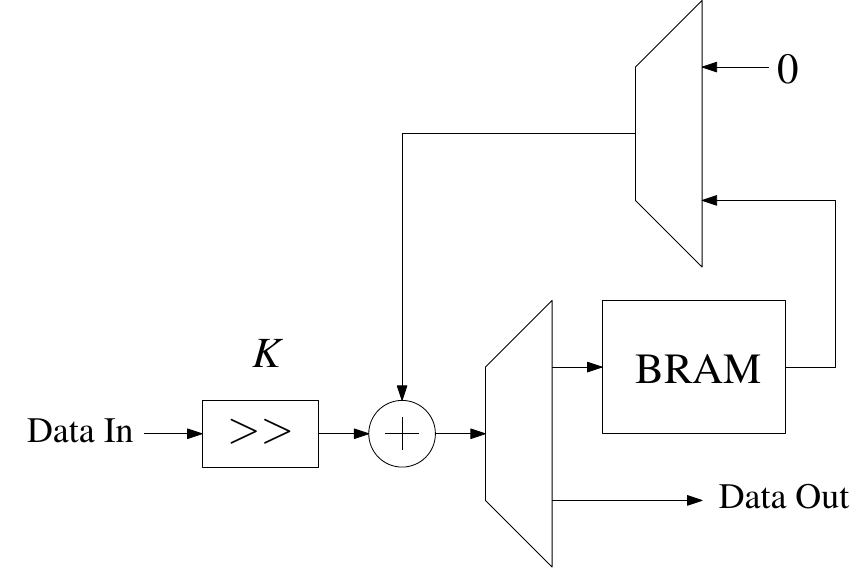}
    \caption{Simplified circuit diagram of the averager within the SounderRx block.}
    \label{fig:circuit}
\end{figure}

The channel sounding block is responsible for discarding samples, block-averaging the received waveforms, and controlling the \ac{rf} switches. The respective samples are discarded by a logic circuit based on a set of counters that can override the valid signal of the AXI-Stream protocol. The non-discarded samples then go to the block averager. Since the data type is complex short, the averaging can efficiently be done by bit-wise right shifting real and imaginary each sample $K$ times and then adding the $i^\mathrm{th}$ sample of each sounding signal. The simplified circuit diagram of the averager is shown in Fig.~\ref{fig:circuit}. The averager consists of an arithmetic shifter, block RAM (BRAM), and an adder. Multiplexers are controlled by a counter for $m=\{1,2,\ldots,M\}$. 1) If $m = 1$: Samples are added with zero and saved in the BRAM. 2) $1 < m < M$: The $i$-th sample comes in and the $i$-th sample is fetched out of the memory, and their sum is stored in the memory. 3) $m=M$: The $i$-th sample comes in, and the $i$-th sample is fetched from the memory. The sum of those samples is presented at the output of the block averager. In the special case of $m = M = 1$, the $i$-th sample is presented directly on the output of the averager.

Due to the high sample rate, four samples are being processed per clock cycle, resulting in a memory width of 128 bits. 
While the sample counter for the averager is enabled by the valid signal of the incoming samples, the counters and logic that control the RF switches (both for the external and internal Tx/Rx switches) are enabled by the \texttt{rx\_running} or \texttt{tx\_running} logic signal to ensure the correct timing. These signals are the main on/off signals for the transmitter and receiver chain.
The clock driving this RFNoC block is \texttt{clk\_radio}, of which the frequency is equal to 125\,MHz.

To store sample data and configure the USRP, a C++ application has been developed that utilizes the UHD library (Release v4.2.0.1), containing all the essential software to control the USRPs. An additional C++ library was also developed to set the parameters ($L$, $P$, $K$, $M$, $R$, $\mathbf{n}_\mathrm{T}$, $\mathbf{n}_\mathrm{R}$) belonging to the RFNoC block.

\subsection{Synchronization}
The USRPs are synchronized via a \ac{pps} signal for time synchronization and a 10 MHz reference signal for frequency synchronization. Through the main application on the host computer, the synchronization source can be set to either a \ac{gps} disciplined oscillator or an external clock (e.g., Rubidium clock).
The channel sounder is designed to start on the positive flank of a \ac{pps}. To eliminate the requirement that two or more USRPs start at the same \ac{pps} flank, the repetition time $T_\mathrm{rep}$ must be chosen in such a way that at each PPS flank a sounding signal starts transmitting, e.g., $T_\mathrm{rep}$~=~50 ms. In this way, multiple USRPs can start at different \ac{pps} flanks; hence, no communication is needed between them. Nevertheless, the USRPs can also start on the same \ac{pps} flank via the \ac{gps} time when the system is synchronized with \ac{gps}.

\subsection{Antenna Design}

\begin{figure}
    \centering
    \includegraphics[width=0.8\columnwidth]{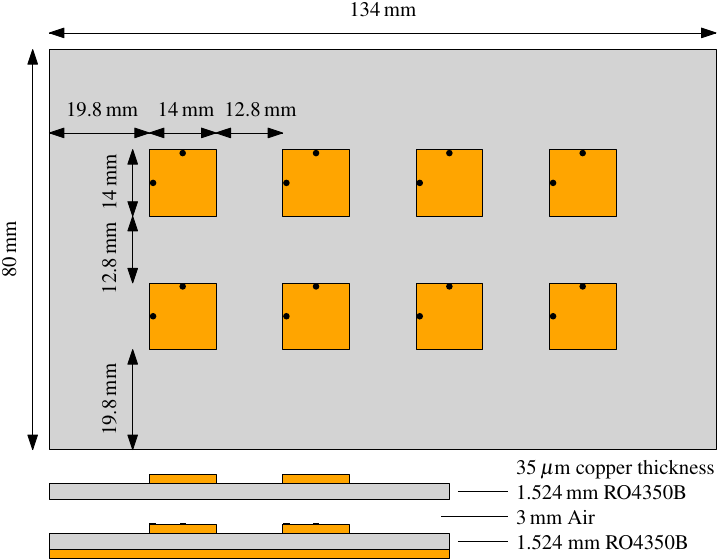}
    \caption{Dimensions of the 16-element antenna array.}
    \label{fig:antenna_dim}
\end{figure}

\begin{figure}
    \centering
    \includegraphics[width=0.8\columnwidth]{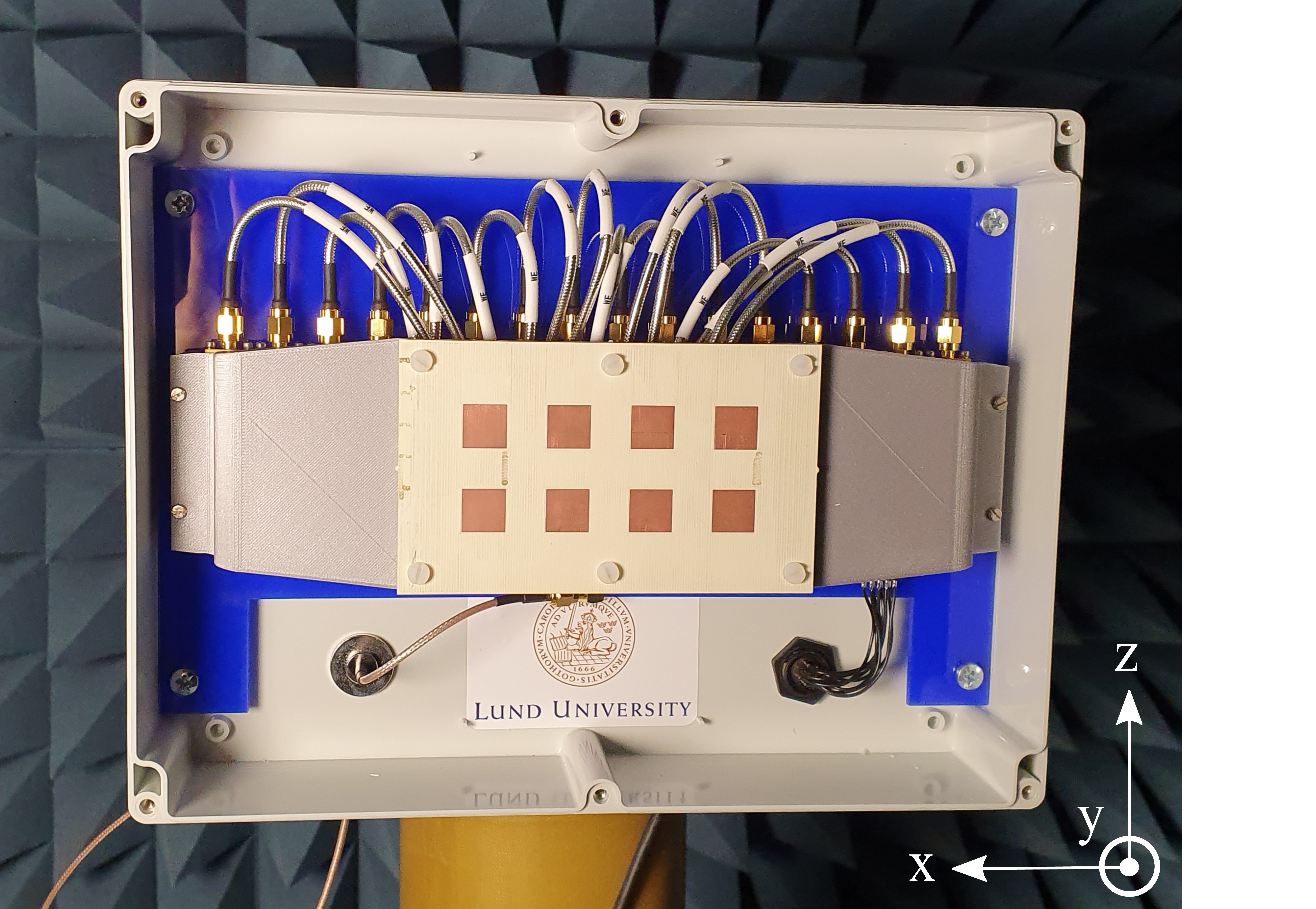}
    \caption{Photograph of one antenna panel in the anechoic chamber.}
    \label{fig:panel}
\end{figure}

\begin{figure}
    \centering
    \input{antenna2d.pgf}
    \caption{Full measured radiation pattern of vertically-polarized antenna element 7 on panel 0 at 5.675\,GHz. The plot shows a typical pattern for a patch antenna. The peak gain value, \textit{including} cable and RF switch losses is -0.4\,dBi. }
    \label{fig:ant1}
\end{figure}
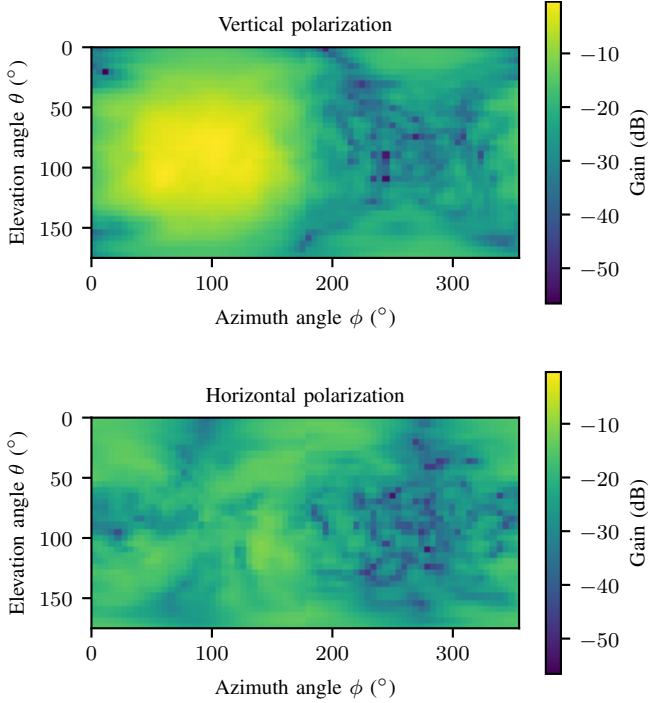

Our implementation comprises eight mini-circuit USB-1SP16T-83H RF switches. For each \ac{rf} switch, we designed a rectangular $2 \times 4$ antenna array with each element having two ports for horizontal and vertical polarization. The geometric centers of the antennas are separated by 26.7\,mm $\approx 0.5 \lambda$. Other dimensions can be found in Fig.~\ref{fig:antenna_dim}. Each element of the array is a proximity-coupled stacked patch antenna and is designed to operate in the 5.0-6.0\,GHz frequency range (S11 $< -10$\,dB). The antenna is made with Rogers Corp. RO4350B laminate. The antennas are mounted on top of the RF switch inside a plastic box, forming one antenna panel. A photograph of one antenna panel is shown in Fig.~\ref{fig:panel}. All panels have been characterized in an anechoic chamber, which is also needed to perform high-resolution angular estimation of multipath components. 
Fig.~\ref{fig:ant1} presents the complete radiation pattern for antenna element 7 of a panel at 5.675\,GHz for both polarizations, showing a typical pattern of a patch antenna. It should be noted that the gain given in the figures includes switch losses and internal cable losses.

\subsection{Final Specifications, Performance and Link Budget}
We summarize the channel sounder specifications in Table~\ref{tab:specs}. Typical parameters for three scenarios and their corresponding performance metrics in are presented in Table~\ref{tab:param}. As shown, the channel sounder can be tailored for wide variety of measurement scenarios and different application requirements. While Scenarios 1 and 2 are further discussed in Section~\ref{sec:verif}, more information about Scenario 3 can be found in \cite{Sandra2023}.

While most of the calculations are straight-forward, special attention should be paid to the calculation of the maximum Doppler frequency. 
From the sampling theorem, it is known that the maximum Doppler frequency of the channel should be smaller than half of the sampling frequency. 
In our channel sounder three frequencies could be regarded as the sampling frequency: \textit{i)} the snapshot rate or the repetition frequency, which has historically been used for the maximum Doppler frequency calculation for channel sounders \cite{book_molisch}, \textit{ii)} the frequency at which each antenna combination is measured \cite{Cai2023}, applied in systems with a large amount of switched antennas, \textit{iii)} the baseband sampling frequency $f_\mathrm{s}$, the fundamental limit. 
For cases \textit{ii)} and \textit{iii)} the channel is not continuously measured but rather in bursts. 
If the Doppler frequency is much lower than the sampling frequencies, it might not be possible to estimate the Doppler frequency because the phase difference between the samples is much smaller than the noise component of the signal. 
Hence, the maximum Doppler frequency should be carefully calculated based on the configuration of the channel sounder and the application scenario. In addition, the average functionality of our channel sounder poses an extra limitation on the Doppler frequency. 
If we repeat the same waveform $\textbf{s} \in \mathbb{C}^{L\times1}$ $M$ times and virtually introduce a Doppler frequency offset $\nu$, the averaged waveform $\textbf{y}  \in \mathbb{C}^{L\times1}$ becomes:
\begin{align}        
    \mathbf{y}(i) &= \frac{1}{2^K} \sum^{M}_{m=1} \mathbf{s}(i) \exp(\mathrm{j}2\pi\nu(mLT_\mathrm{s}+iT_\mathrm{s})) \nonumber \\
    &= \frac{1}{2^K}\,\mathbf{s}(i) \exp(\mathrm{j}2\pi\nu iT_\mathrm{s}) \sum^{M}_{m=1} \exp(\mathrm{j}2\pi\nu mLT_\mathrm{s}). \label{eq:avg_doppler}
\end{align}
The summation in \eqref{eq:avg_doppler} is a sum of phasors. If the phases of these phasors do not align well, e.g. due to a high Doppler frequency, this sum can attenuate the signal drastically. We consider 
\begin{equation}
    \nu_{\mathrm{max}} \leq \dfrac{0.1 f_\mathrm{s}}{ML}
\end{equation}
to be an acceptable limit.
In addition, it must be noted that the processing gain does not include the beamforming gain because it depends on the \ac{aod} and \ac{aod}.


\begin{table}
\small
\centering
\caption{Final specifications of the channel sounder \label{tab:specs}}
\begin{tabular}{@{}ll@{}}
\toprule
\textbf{Specification} & \textbf{Value} \\ \midrule
Type & Hybrid-switched \\
Operating frequency & 5$-$6\,GHz\\
Bandwidth & 400\,MHz\\
Number of antenna panels & 8 \\
Panel array configuration & 2$\times$4 dual polarized\\
Panel azimuth/elevation 10\,dB beamwidths & 120/120 degrees \\ 
Number of standalone antennas & 4 \\
Maximum number of antenna combinations & 7656 \\
Maximum RF switch frequency & 50\,kHz\\
Typical snapshot rate & 10$-$200\,Hz \\
Typical output power & 18\,dBm \\
ADC/DAC depth & 12/14\,bits \\ 
\bottomrule
\end{tabular}
\end{table}

\begin{table*}[]
    \centering
        \caption{Typical configuration of the channel sounder and corresponding performance metrics}
    \label{tab:param}
    \begin{tabular}{llccc}
    \toprule
    \textbf{Description} & \textbf{Parameter} &  \textbf{Scenario 1} & \textbf{Scenario 2} & \textbf{Scenario 3}\\ \midrule
         Center frequency & $f_\mathrm{c}$ & 5.675\,GHz& 5.675\,GHz& 5.6\,GHz\\ 
         Sample frequency & $f_\mathrm{s}$  & 500\,Msps & 500\,Msps & 500\,Msps \\   
         Sounding waveform length & $L$ & 1024 & 1024 & 8192 \\
         Number of tones & $F$  & 819 & 819 & 4095 \\
         Number of waveforms to sum in averager & $M$ & 8 & 8 & 64\\
         Division of $2^K$ in averager & $K$ & 3 & 3 & 2 \\
         Skipped samples & $P$ & 9216 & 9216 & 49152\\
         Skipped samples & $R$& 2221472 & 14348416 & 1975712\\ 
         Number of RF chains & $N_\mathrm{RF}$ & 9 & 8 & 2 \\ 
         Total number of transmit antennas & $\mathrm{sum}\{\textbf{n}_\mathrm{T}\}$ & 1 & 128 &1\\
         Maximum amount of panel antennas & $\mathrm{max}\{\textbf{n}_\mathrm{R}\}$ & 16 & 16 &1\\
        \midrule
        \textbf{Metric} & \textbf{Equation} &   & & \\ \midrule
        Bandwidth & $f_\mathrm{s} F / L$ & 400\,MHz & 400\,MHz& 250\,MHz \\[1ex]
        Snapshot rate $f_\mathrm{rep}=1/T_\mathrm{rep}$ & $f_\mathrm{s}/(\mathrm{sum}\{\textbf{n}_\mathrm{T}\} \mathrm{max}\{\textbf{n}_\mathrm{R}\}(P+ML)+R)$ & 200\,Hz& 10\,Hz& 200\,Hz\\[1ex]
        Coherence time & $\mathrm{sum}\{\textbf{n}_\mathrm{T}\} \mathrm{max}\{\textbf{n}_\mathrm{R}\}(P+ML)/f_\mathrm{s}$&  557\,$\mu$s & 71\,ms & 2\,ms\\[1ex]
        Antenna combinations & see \eqref{eq:nantcomb} & 128 & 7168 & 1\\[1ex]
        Max. delay spread & $L/f_\mathrm{s}$& 2\,$\mu$s & 2\,$\mu$s& 16\,$\mu$s \\[1ex]
        Max. Doppler frequency & $\mathrm{min}\left(f_\mathrm{rep}\,\text{or}\,\dfrac{f_\mathrm{s}}{P+ML}, \dfrac{0.1 f_\mathrm{s}}{ML}\right)$ & 29\,kHz & 29\,kHz & 100 Hz\\[1ex]
        Processing gain (w/o beamforming gain) & $10\,\log_{10}(ML)$ & 39\,dB & 39\,dB & 57\,dB\\[1ex]
        Sensitivity level (20\,dB SNR)\footnote{The beamforming is not taken into account as this is depending on the angle of departure and angle of arrival} &$-147.5+10\,\log_{10}(f_\mathrm{s}FM)$& -100\,dBm&-100\,dBm&-118\,dBm\\[1ex]
        Data rate USRP to host (w/o overhead) & $\dfrac{4 f_\mathrm{s} L\,\mathrm{sum}\{\textbf{n}_\mathrm{T}\} \mathrm{max}\{\textbf{n}_\mathrm{R}\} N_\mathrm{RF,USRP}}{\mathrm{sum}\{\textbf{n}_\mathrm{T}\} \mathrm{max}\{\textbf{n}_\mathrm{R}\}(P+ML)+R}$ & 236 MB/s
        & 671 MB/s & 3.2 MB/s \\
        \bottomrule
    \end{tabular}
\end{table*}

\begin{figure}
    \centering
    \includegraphics[width=0.95\columnwidth]{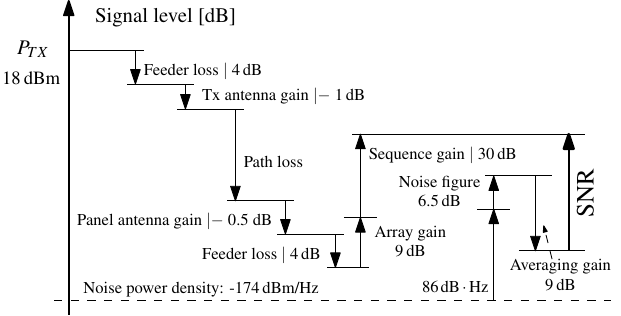}
    \caption{Link-budget analysis of a standalone antenna as transmitter to one antenna panel as receiver. }
    \label{fig:linkbudget}
\end{figure}

A link budget analysis is presented in Fig.~\ref{fig:linkbudget} of a single antenna as transmitter and a panel as receiver. Note that the noise figure is dependent on the receiver gain setting on the USRP and temperature-variant. The noise figure given in the figure is a typical value under maximum gain configuration.



\section{Post-processing and calibration} \label{sect3}
The output data of the channel data can be represented as a six-dimensional tensor $\textbf{D}(s, p_\mathrm{T}, p_\mathrm{R}, m_\mathrm{T}, m_\mathrm{R}, n)$, where $s$ is the snapshot index, $n_\mathrm{T}$ the transmit chain index, $n_\mathrm{R}$ the receive chain index, $m_\mathrm{T}$ the transmit antenna index, $m_\mathrm{R}$ the receive antenna index, and $i$ the sample index. Further data processing is done by performing a Discrete Fourier Transform over $i$ to obtain $\textbf{Y}(s, p_\mathrm{T}, p_\mathrm{R}, m_\mathrm{T}, m_\mathrm{R}, k)$ where $k$ is the frequency index. Hereafter, the frequency indices are selected where a signal was transmitted, i.e. $\textbf{x}(k) \neq 0$. The complete channel response $\textbf{H}$ is then calculated by diving $\textbf{Y}$ by $\textbf{x}$ in the frequency domain:
\begin{align}
    \textbf{H}(s, p_\mathrm{T}, p_\mathrm{R}, m_\mathrm{T}, m_\mathrm{R}, k) = \dfrac{\textbf{Y}(s, p_\mathrm{T}, p_\mathrm{R}, m_\mathrm{T}, m_\mathrm{R}, k)}{\textbf{x}(k)}.
\end{align}

Note that $\textbf{H}$ includes the response of the \ac{rf} chains and the antennas, which is not part of the propagation channel. Although this result could already be used for some simulations, our ultimate goal is to obtain the true propagation characteristics. This can be done by modeling the propagation channel as a superposition of specular \acp{mpc},
which each have the following parameters: delay ($\tau$), Doppler frequency ($\nu$), azimuth and elevation \ac{aod} ($\phi_\mathrm{T}$, $\theta_\mathrm{T}$), azimuth and elevation \ac{aoa} ($\phi_\mathrm{R}$, $\theta_\mathrm{R}$), 
and complex polarimetric path gains ($\boldsymbol{\gamma}$). In addition, it is assumed that these geometrical propagation parameters of \acp{mpc} are unchanged during the observation time of $S$ snapshots. Analytically,
\begin{equation}
\begin{split}
      &\textbf{H}(s, p_\mathrm{T}, p_\mathrm{R}, m_\mathrm{T}, m_\mathrm{R}, k)
     \\= & \sum_{l=1}^L \textbf{a}_{m_\mathrm{R},p_\mathrm{R}}^\mathrm{T}(\phi_{\mathrm{R},l}, \theta_{\mathrm{R},l}, f_k)  
     \begin{pmatrix}
     \gamma_{\mathrm{HH},l} & \gamma_{\mathrm{HV},l}\\
     \gamma_{\mathrm{VH},l}& \gamma_{\mathrm{VV},l}\\
     \end{pmatrix}
\\&\textbf{a}_{m_\mathrm{T},p_\mathrm{T},}(\phi_{\mathrm{T},l}, \theta_{\mathrm{T},l}, f_k)  \textbf{b}_{p_\mathrm{T}, n_\mathrm{R}}(k) \exp(-\mathrm{j}2\pi f_k \tau_l) \\
     &  \exp(\mathrm{j}2 \pi \nu_l t_{s, p_\mathrm{T}, p_\mathrm{R}, m_\mathrm{T}, m_\mathrm{R}}) +\textbf{N} (s, p_\mathrm{T}, p_\mathrm{R}, m_\mathrm{T}, m_\mathrm{R}, f_k),
\end{split}
\end{equation}
where $f_k$ is the absolute frequency value that corresponds with frequency index $k$, $l$ is the index of the \ac{mpc}, $t_{s, p_\mathrm{T}, p_\mathrm{R}, m_\mathrm{T}, m_\mathrm{R}}$ gives the time in seconds when the channel between the respective antenna elements was measured, $\textbf{N}$ gives the measurement noise. Parameters $ \gamma_{\mathrm{HH},l}, \gamma_{\mathrm{HV},l}, \gamma_{\mathrm{VH},l},\gamma_{\mathrm{VV},l}$ denote the horizontal-to-horizontal, horizontal-to-vertical, vertical-to-horizontal, and vertical-to-vertical polarization gains of the $l$-th \ac{mpc}.
Moreover, $\textbf{a}_{m_\mathrm{T},p_\mathrm{T}} \in \mathbb{C}^{2\times1}$ and $\textbf{a}_{m_\mathrm{R},p_\mathrm{R}}\in \mathbb{C}^{2\times1}$ denote the polarimetric antenna responses. These antenna responses are obtained through the effective aperture distribution function (EADF) \cite{Richter2005, Cai2023a} based on the antenna panel calibration data. Furthermore, $\textbf{b}_{p_\mathrm{T}, p_\mathrm{R}}$ contains the frequency response of the sounder without antennas, which can be obtained from a back-to-back measurement with an attenuator. Hence,
\begin{equation}
     \textbf{b}_{p_\mathrm{T}, p_\mathrm{R}}(k) = \dfrac{\textbf{Y}(0, p_\mathrm{T}, p_\mathrm{R}, 0, 0, k)}{\textbf{x}(k) \textbf{s}_{21,\mathrm{att}}(k)},
\end{equation}
where $\textbf{s}_{21,\mathrm{att}} \in \mathbb{C}^{F\times1}$ is the frequency response of the attenuator measured with a VNA. Note that due to the design of the channel sounder, it is not possible to measure the response between the transmit and receive chains on the same channel, i.e. $n_\mathrm{T} \neq n_\mathrm{R}$. 
Measurement of all RF channel combinations can be cumbersome and time consuming. $N_\mathrm{RF}(N_\mathrm{RF}-1)/2$ connections have to be made. To decrease this amount, it is possible to obtain all combinations $\textbf{b}_{p_\mathrm{T}, p_\mathrm{R}}$ by measuring only a set of combinations, which needs $2 (N_\mathrm{RF}-2)+1$ back-to-back connections. Analytically,
\begin{equation}
     \textbf{b}_{p_\mathrm{T}, p_\mathrm{R}}(k) = \dfrac{\textbf{b}_{0, p_\mathrm{R}}(k) \textbf{b}_{p_\mathrm{T}, 1}(k) }{\textbf{b}_{0, 1}(k)}.
\end{equation}
Several approaches exist to estimate the \ac{mpc} parameters $(\tau, \nu, \phi_\mathrm{T}, \theta_\mathrm{T}, \phi_\mathrm{R}, \theta_\mathrm{R}, \boldsymbol{\gamma})$ such as expectation-maximization (EM)\cite{Feder1988}, space-alternating generalized expectation maximization (SAGE) \cite{Fessler1994, Fleury1999}, and RIMAX \cite{Richter2005}. To process our verification measurements in Section~\ref{sec:verif}, we have implemented and applied the SAGE algorithm.



\section{Verification and Measurements \label{sec:verif}} \label{sect4}

\subsection{Scenario 1: Uplink Measurement \label{sec:exp1}}
\begin{figure}
    \centering
    \includegraphics[width=0.9\columnwidth]{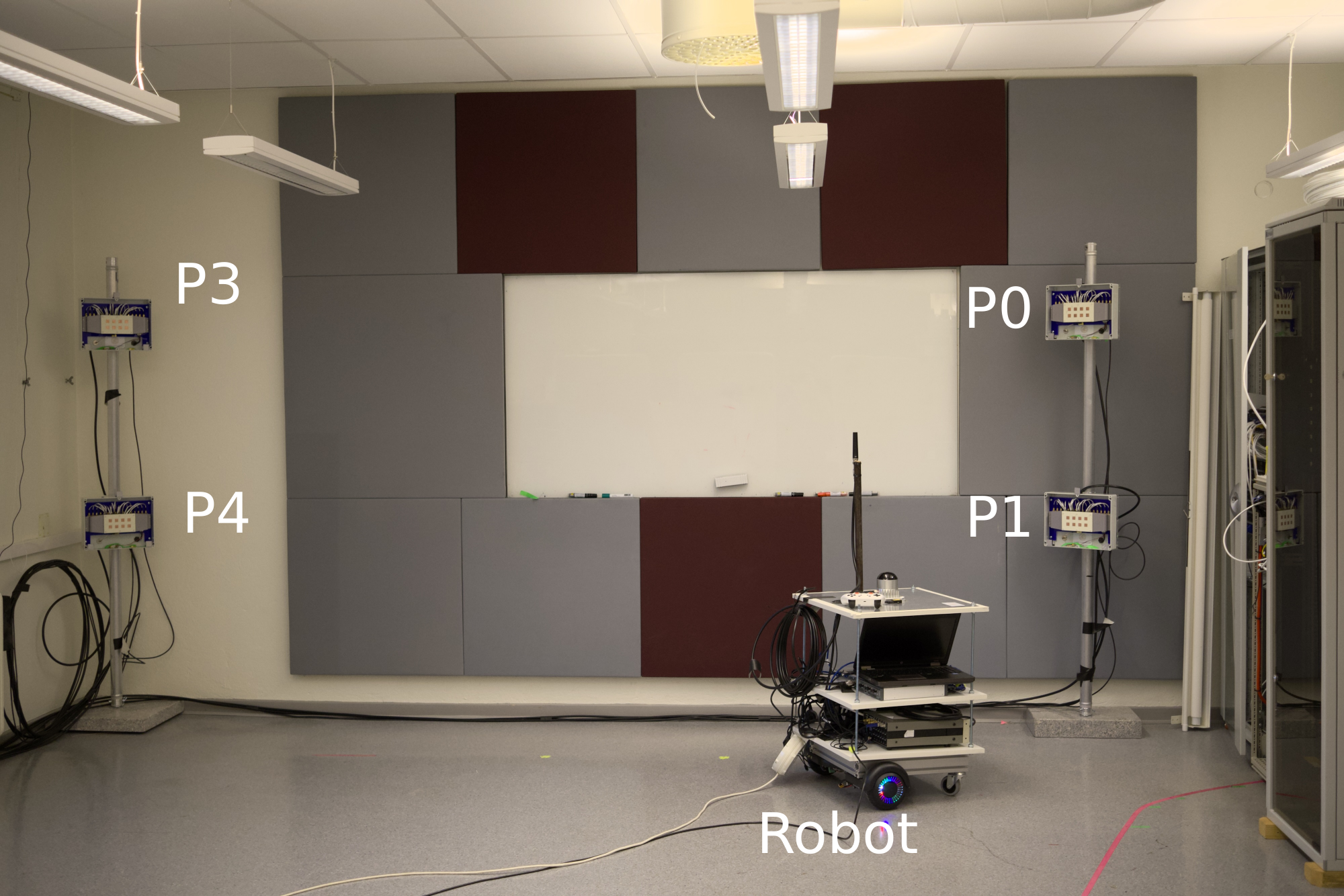}
    \caption{Photograph of the measurement setup and environment. Four of the eight panels (P3, P4, P0, P1) are visible: two on the left and two on the right. The robot is located in front of the panels.}
    \label{fig:setup}
\end{figure}

\begin{figure}
    \centering
    \resizebox{200pt}{!}{
    \input{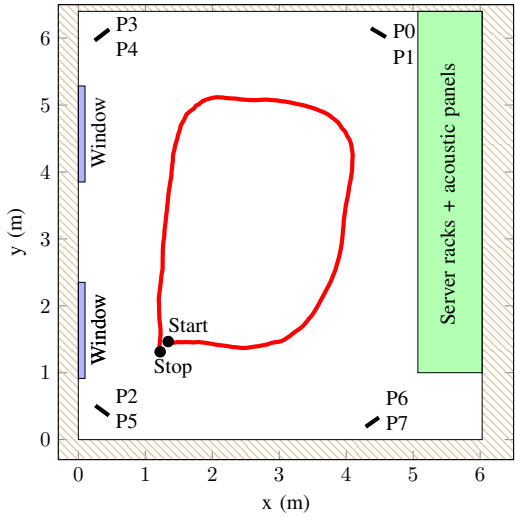}
    }
    \caption{Floor plan of the measurement environment and the trajectory for the uplink measurement given in red.}
    \label{fig:setup_floorplan}
\end{figure}

In order to verify the functionality of the channel sounder, we have performed a measurement of the propagation channel at 5.675\,GHz between a single antenna robot and eight distributed panels in an indoor laboratory environment with a bandwidth of 400\,MHz. The laboratory room, depicted in Fig.~\ref{fig:setup}, measures approximately 5 meters by 12 meters, of which an area of 5 meters by 6 meters was used for the measurements. Eight panels were distributed in the room in groups of two, as illustrated in Fig.~\ref{fig:setup_floorplan}. Each group of two panels was vertically stacked at heights of 0.92\,m and 1.84\,m. Then each group of four panels was connected to a USRP X410 through 10\,m long LMR400 RF cables and custom-made cables for switch control. The two USRPs were connected to an industrial computer through a link 100\,GbE and are synchronized through one Rubidium clock.

\begin{figure}
    \centering
    \input{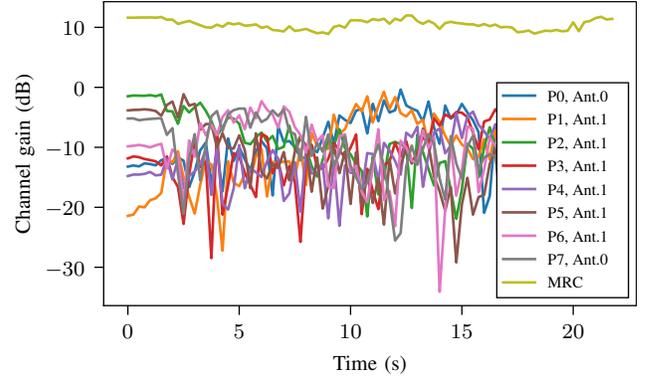}
    \caption{Narrowband channel gain at 5.675\,GHz for a single vertically-polarized antenna per panel and the result after performing maximum ratio combining (MRC) of all 128 antennas.}
    \label{fig:mrc_result}    
\end{figure}

\begin{figure*}
    \centering
    \includegraphics[width=0.9\textwidth]{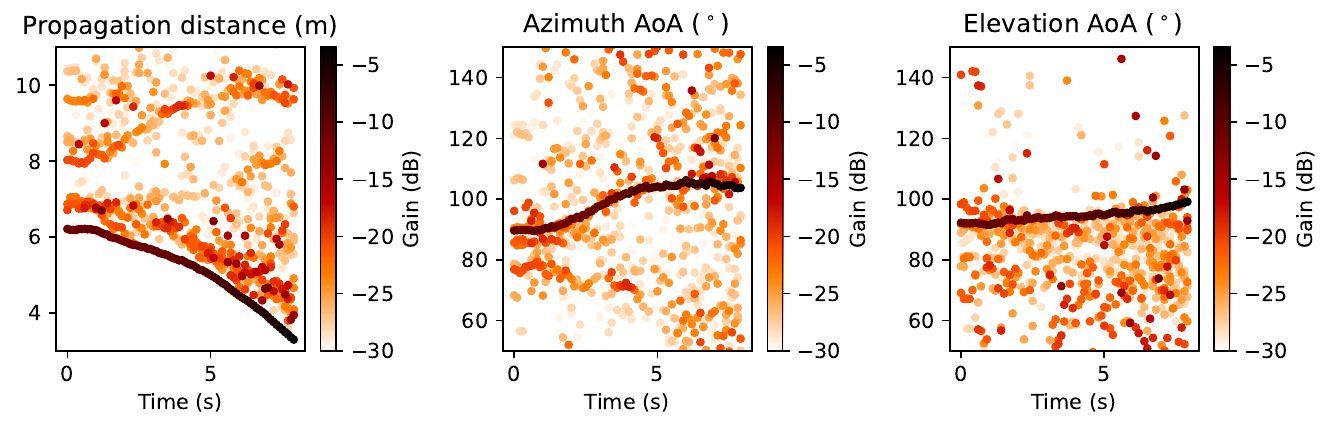}
    \caption{High-resolution parameter estimates of the \acp{mpc} seen from P0 using the SAGE algorithm for the first eight seconds of the uplink channel measurement. From left to right: propagation distance, azimuth AoA, elevation AoA. }
    \label{fig:sage_result}    
\end{figure*}

The robot is equipped with a USRP X410, a wideband dipole antenna, two regular laptops, a Rubidium reference clock, uninterruptible power supply (UPS), a \ac{lidar} sensor and an \ac{imu}. A laptop was responsible for controlling the USRP and the movement of the robot. The other laptop was used to collect data from the \ac{lidar} and \ac{imu}, which provide ground truth information about the location of the robot and the panels in the room via a \ac{slam} algorithm. Reflective tape was applied to the panels to enhance their visibility in the \ac{lidar} scan. The robot is an in-house design based on a hoverboard and is controlled via a game controller over Bluetooth. A detailed list of primary equipment can be found in Table~\ref{tab:device}. 

\begin{table}
\centering
\caption{Primary equipment\label{tab:device}}
\begin{tabular}{@{}ccl@{}}
\toprule
\textbf{Device} & \textbf{Amount} & \textbf{Model} \\ \midrule
Software-defined radio & 3 & NI USRP X410 \\
Industrial computer &1 & Advanctech MIC-770 V2\\
Dipole antenna & 1 & Taoglas TU.60.3H31 \\
Rubidium clock & 2 & SRS FS725 \& SRS FS740 \\
Lidar & 1 & Ouster OSDome (128 lines) \\
IMU & 1 & Microstrain 3DM-GX5-25 (AHRS)\\
RF switches (panel) & 8 & Mini-Circuits USB-1SP16T-83H \\ 

\bottomrule
\end{tabular}
\end{table}

During the measurement, the robot followed a round trajectory, maintaining a \ac{los} connection with all panels, as illustrated in Fig.~\ref{fig:setup_floorplan}. The channel sounder configuration during this measurement can be found in Table~\ref{tab:param}.


Fig.~\ref{fig:mrc_result} presents the narrowband channel gains during the measurement for several vertically polarized antenna elements at different panels, and the channel gain after applying maximum ratio combining (MRC) for all 128 antennas. It can be observed that with MRC the overall channel gain increased. The deep small-scale fading experienced by every single antenna has also been effectively mitigated. This demonstrates the potential for the reliability increase with distributed massive \ac{mimo} systems. 
Furthermore, the SAGE algorithm is applied to extract the parameters of the \acp{mpc} observed by panel 0 are depicted in Fig.~\ref{fig:sage_result} for the initial eight seconds of measurement. The \ac{los} component is readily apparent.
In the propagation distance plot, we discern two additional tracks of \acp{mpc}. The track closest to the \ac{los} emanates from reflections off the ceiling and ground. The other track, exhibiting increasing propagation distance, is most likely a reflection off one of the side walls of the room. These two tracks are also visible in the azimuth plot: the ground/ceiling reflections align closest to the \ac{los} azimuth angle, while the other track ranges from around 77 degrees to about 70 degrees. Examining the elevation plot, we observe that \acp{mpc} originate primarily from the ceiling, attributable to the lighting infrastructure, causing significant scattering.

\subsection{Scenario 2: Sensing Measurement}
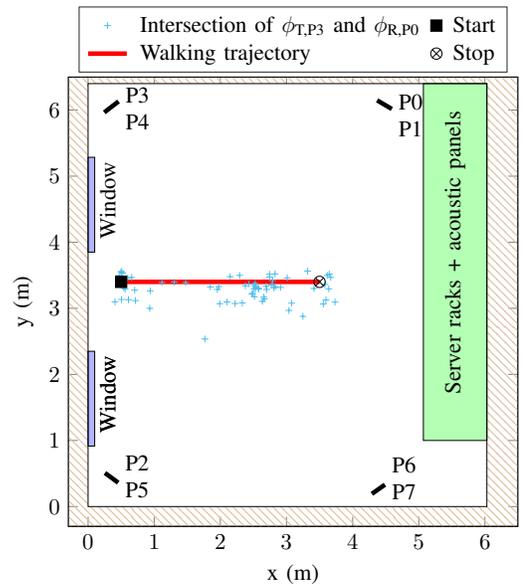
\begin{figure}
    \centering
    \resizebox{200pt}{!}{
    \input{scenarios2.tex}
    }
    \caption{Floor plan of the measurement environment. The walking trajectory of a person is given by the red line. The blue crosses are the intersections of the AoD from P3 and AoA of P1, after selecting a parameter subspace.}
    \label{fig:setup_floorplan_sens}
\end{figure}

\begin{figure*}
    \centering
    \includegraphics[width=0.9\textwidth]{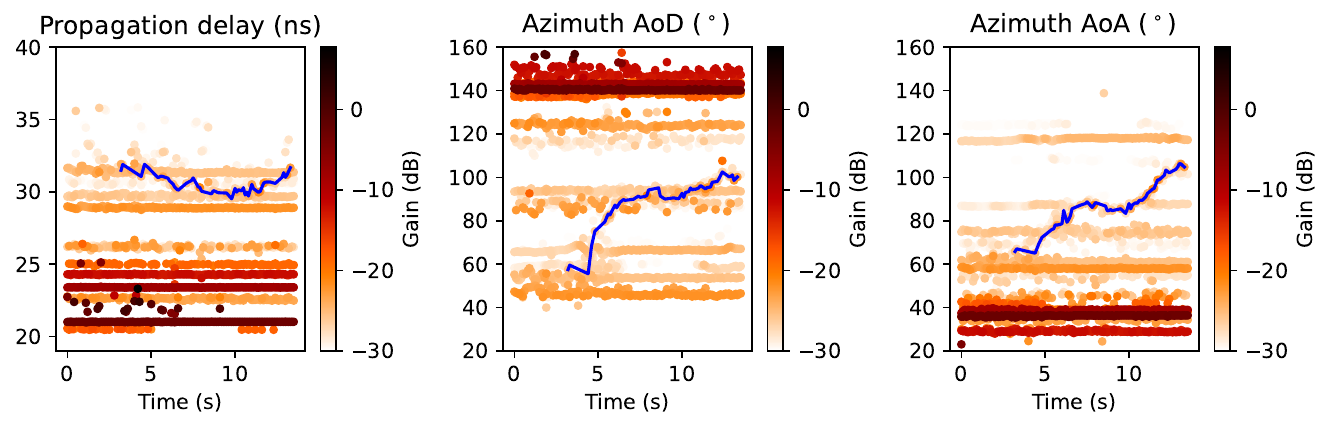}
    \caption{High-resolution parameter estimates of the \acp{mpc} between P1 (Rx) and P3 (Tx) using the SAGE algorithm during measurement scenario 2. From left
to right: propagation delay, azimuth AoD, azimuth AoA}
    \label{fig:sage_result_sens}    
\end{figure*}

To demonstrate the sensing capability of our channel sounder, we have performed an additional measurement in the same environment as in Section~\ref{sec:exp1}. 
In contrast with the previous measurement, all eight panels are now configured as both transmitting and receiving. 
This configuration allows us to estimate the AoD from each panel, which can be useful for sensing purposes. The channel sounder configuration can be found in Tab.~\ref{tab:param} under Scenario 2. 
The goal of this measurement was to test the capability of sensing a person with our channel sounder, based on the estimated parameters of the \acp{mpc}.
During the measurement, a member of research group walked on the red line shown in Fig.~\ref{fig:setup_floorplan_sens} within a time of approximately 10 seconds. The person stood still at the beginning and at the middle of the walk. 
The SAGE estimation results of the propagation delay ($\tau$), azimuth AoD ($\phi_\mathrm{T}$) and AoA ($\phi_\mathrm{R}$) are presented in Fig.~\ref{fig:sage_result_sens}. The figures show that most of the estimates are not time-dependent because the panels and the environment are static, except for the walking person. However, we are interested in the \acp{mpc} that change with time because these could be related to the moving person. At 5 seconds in the \ac{aod} plot, a change in the \ac{aod} can be observed. To get a better observation of the dynamic \acp{mpc}, we have manually selected a parameter subspace $\{\tau = [29\,\mathrm{ns}, 32\,\mathrm{ns}],\phi_\mathrm{R} = [65^\circ,110^\circ], \theta_\mathrm{T} = [70^\circ,180^\circ]\}$. The \acp{mpc} in this subspace are marked with the blue line Fig.~\ref{fig:sage_result_sens}. To verify the \acp{mpc} in this subspace are linked to the moving person, we calculated the intersection between the AoD and the AoA, indicated with the blue crosses in Fig.~\ref{fig:setup_floorplan_sens}. Because intersection points are in close proximity of the red line, we can confident that the \acp{mpc} are connected to the movement of the person.

%
\section{Conclusions} \label{sect5}
In this paper we have presented the design, implementation and verification of a channel sounder for distributed massive \ac{mimo} based on open-source software and off-the-shelf electronic components.
Our design utilized multiple NI USRP X410 and radio frequency switches as core components. 
By means of a flexible design, customized \ac{fpga} code, and integration of multiple parallel transceiver chains and \ac{rf} switches, we were able to scale up the number of antennas in a cost-efficient way and without a significant compromise in dynamic capability.
The combination of the high number of antennas, wide bandwidth and multi-link capability makes this design a valuable asset for conducting cutting-edge research within areas such as distributed massive \ac{mimo} and integrated sensing and communication. 
This potential was demonstrated through two measurement campaigns in our laboratory, one distributed massive \ac{mimo} uplink measurement and one passive sensing measurement.

\section*{Acknowledgement}
The authors would like to thank Yingjie Xu for his assistance during the measurements and Guoda Tian for proofreading the manuscript.



\ifCLASSOPTIONcaptionsoff
  \newpage
\fi
\bibliographystyle{IEEEtran}
\bibliography{IEEEabrv,jabref, references}





\end{document}

%% file: antenna2d.pgf
\begingroup%
\makeatletter%
\begin{pgfpicture}%
\pgfpathrectangle{\pgfpointorigin}{\pgfqpoint{3.367874in}{3.653120in}}%
\pgfusepath{use as bounding box, clip}%
\begin{pgfscope}%
\pgfsetbuttcap%
\pgfsetmiterjoin%
\definecolor{currentfill}{rgb}{1.000000,1.000000,1.000000}%
\pgfsetfillcolor{currentfill}%
\pgfsetlinewidth{0.000000pt}%
\definecolor{currentstroke}{rgb}{1.000000,1.000000,1.000000}%
\pgfsetstrokecolor{currentstroke}%
\pgfsetdash{}{0pt}%
\pgfpathmoveto{\pgfqpoint{-0.000000in}{0.000000in}}%
\pgfpathlineto{\pgfqpoint{3.367874in}{0.000000in}}%
\pgfpathlineto{\pgfqpoint{3.367874in}{3.653120in}}%
\pgfpathlineto{\pgfqpoint{-0.000000in}{3.653120in}}%
\pgfpathlineto{\pgfqpoint{-0.000000in}{0.000000in}}%
\pgfpathclose%
\pgfusepath{fill}%
\end{pgfscope}%
\begin{pgfscope}%
\pgfsetbuttcap%
\pgfsetmiterjoin%
\definecolor{currentfill}{rgb}{1.000000,1.000000,1.000000}%
\pgfsetfillcolor{currentfill}%
\pgfsetlinewidth{0.000000pt}%
\definecolor{currentstroke}{rgb}{0.000000,0.000000,0.000000}%
\pgfsetstrokecolor{currentstroke}%
\pgfsetstrokeopacity{0.000000}%
\pgfsetdash{}{0pt}%
\pgfpathmoveto{\pgfqpoint{0.440443in}{2.310888in}}%
\pgfpathlineto{\pgfqpoint{2.678739in}{2.310888in}}%
\pgfpathlineto{\pgfqpoint{2.678739in}{3.414273in}}%
\pgfpathlineto{\pgfqpoint{0.440443in}{3.414273in}}%
\pgfpathlineto{\pgfqpoint{0.440443in}{2.310888in}}%
\pgfpathclose%
\pgfusepath{fill}%
\end{pgfscope}%
\begin{pgfscope}%
\pgfpathrectangle{\pgfqpoint{0.440443in}{2.310888in}}{\pgfqpoint{2.238295in}{1.103385in}}%
\pgfusepath{clip}%
\pgfsys@transformshift{0.440443in}{2.310888in}%
\pgftext[left,bottom]{\includegraphics[interpolate=true,width=2.240000in,height=1.110000in]{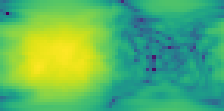}}%
\end{pgfscope}%
\begin{pgfscope}%
\pgfsetbuttcap%
\pgfsetroundjoin%
\definecolor{currentfill}{rgb}{0.000000,0.000000,0.000000}%
\pgfsetfillcolor{currentfill}%
\pgfsetlinewidth{0.803000pt}%
\definecolor{currentstroke}{rgb}{0.000000,0.000000,0.000000}%
\pgfsetstrokecolor{currentstroke}%
\pgfsetdash{}{0pt}%
\pgfsys@defobject{currentmarker}{\pgfqpoint{0.000000in}{-0.048611in}}{\pgfqpoint{0.000000in}{0.000000in}}{%
\pgfpathmoveto{\pgfqpoint{0.000000in}{0.000000in}}%
\pgfpathlineto{\pgfqpoint{0.000000in}{-0.048611in}}%
\pgfusepath{stroke,fill}%
}%
\begin{pgfscope}%
\pgfsys@transformshift{0.440443in}{2.310888in}%
\pgfsys@useobject{currentmarker}{}%
\end{pgfscope}%
\end{pgfscope}%
\begin{pgfscope}%
\definecolor{textcolor}{rgb}{0.000000,0.000000,0.000000}%
\pgfsetstrokecolor{textcolor}%
\pgfsetfillcolor{textcolor}%
\pgftext[x=0.440443in,y=2.213666in,,top]{\color{textcolor}\rmfamily\fontsize{8.000000}{9.600000}\selectfont \(\displaystyle {0}\)}%
\end{pgfscope}%
\begin{pgfscope}%
\pgfsetbuttcap%
\pgfsetroundjoin%
\definecolor{currentfill}{rgb}{0.000000,0.000000,0.000000}%
\pgfsetfillcolor{currentfill}%
\pgfsetlinewidth{0.803000pt}%
\definecolor{currentstroke}{rgb}{0.000000,0.000000,0.000000}%
\pgfsetstrokecolor{currentstroke}%
\pgfsetdash{}{0pt}%
\pgfsys@defobject{currentmarker}{\pgfqpoint{0.000000in}{-0.048611in}}{\pgfqpoint{0.000000in}{0.000000in}}{%
\pgfpathmoveto{\pgfqpoint{0.000000in}{0.000000in}}%
\pgfpathlineto{\pgfqpoint{0.000000in}{-0.048611in}}%
\pgfusepath{stroke,fill}%
}%
\begin{pgfscope}%
\pgfsys@transformshift{1.070949in}{2.310888in}%
\pgfsys@useobject{currentmarker}{}%
\end{pgfscope}%
\end{pgfscope}%
\begin{pgfscope}%
\definecolor{textcolor}{rgb}{0.000000,0.000000,0.000000}%
\pgfsetstrokecolor{textcolor}%
\pgfsetfillcolor{textcolor}%
\pgftext[x=1.070949in,y=2.213666in,,top]{\color{textcolor}\rmfamily\fontsize{8.000000}{9.600000}\selectfont \(\displaystyle {100}\)}%
\end{pgfscope}%
\begin{pgfscope}%
\pgfsetbuttcap%
\pgfsetroundjoin%
\definecolor{currentfill}{rgb}{0.000000,0.000000,0.000000}%
\pgfsetfillcolor{currentfill}%
\pgfsetlinewidth{0.803000pt}%
\definecolor{currentstroke}{rgb}{0.000000,0.000000,0.000000}%
\pgfsetstrokecolor{currentstroke}%
\pgfsetdash{}{0pt}%
\pgfsys@defobject{currentmarker}{\pgfqpoint{0.000000in}{-0.048611in}}{\pgfqpoint{0.000000in}{0.000000in}}{%
\pgfpathmoveto{\pgfqpoint{0.000000in}{0.000000in}}%
\pgfpathlineto{\pgfqpoint{0.000000in}{-0.048611in}}%
\pgfusepath{stroke,fill}%
}%
\begin{pgfscope}%
\pgfsys@transformshift{1.701455in}{2.310888in}%
\pgfsys@useobject{currentmarker}{}%
\end{pgfscope}%
\end{pgfscope}%
\begin{pgfscope}%
\definecolor{textcolor}{rgb}{0.000000,0.000000,0.000000}%
\pgfsetstrokecolor{textcolor}%
\pgfsetfillcolor{textcolor}%
\pgftext[x=1.701455in,y=2.213666in,,top]{\color{textcolor}\rmfamily\fontsize{8.000000}{9.600000}\selectfont \(\displaystyle {200}\)}%
\end{pgfscope}%
\begin{pgfscope}%
\pgfsetbuttcap%
\pgfsetroundjoin%
\definecolor{currentfill}{rgb}{0.000000,0.000000,0.000000}%
\pgfsetfillcolor{currentfill}%
\pgfsetlinewidth{0.803000pt}%
\definecolor{currentstroke}{rgb}{0.000000,0.000000,0.000000}%
\pgfsetstrokecolor{currentstroke}%
\pgfsetdash{}{0pt}%
\pgfsys@defobject{currentmarker}{\pgfqpoint{0.000000in}{-0.048611in}}{\pgfqpoint{0.000000in}{0.000000in}}{%
\pgfpathmoveto{\pgfqpoint{0.000000in}{0.000000in}}%
\pgfpathlineto{\pgfqpoint{0.000000in}{-0.048611in}}%
\pgfusepath{stroke,fill}%
}%
\begin{pgfscope}%
\pgfsys@transformshift{2.331960in}{2.310888in}%
\pgfsys@useobject{currentmarker}{}%
\end{pgfscope}%
\end{pgfscope}%
\begin{pgfscope}%
\definecolor{textcolor}{rgb}{0.000000,0.000000,0.000000}%
\pgfsetstrokecolor{textcolor}%
\pgfsetfillcolor{textcolor}%
\pgftext[x=2.331960in,y=2.213666in,,top]{\color{textcolor}\rmfamily\fontsize{8.000000}{9.600000}\selectfont \(\displaystyle {300}\)}%
\end{pgfscope}%
\begin{pgfscope}%
\definecolor{textcolor}{rgb}{0.000000,0.000000,0.000000}%
\pgfsetstrokecolor{textcolor}%
\pgfsetfillcolor{textcolor}%
\pgftext[x=1.559591in,y=2.050580in,,top]{\color{textcolor}\rmfamily\fontsize{8.000000}{9.600000}\selectfont Azimuth angle \(\displaystyle \phi\) (\(\displaystyle ^\circ\))}%
\end{pgfscope}%
\begin{pgfscope}%
\pgfsetbuttcap%
\pgfsetroundjoin%
\definecolor{currentfill}{rgb}{0.000000,0.000000,0.000000}%
\pgfsetfillcolor{currentfill}%
\pgfsetlinewidth{0.803000pt}%
\definecolor{currentstroke}{rgb}{0.000000,0.000000,0.000000}%
\pgfsetstrokecolor{currentstroke}%
\pgfsetdash{}{0pt}%
\pgfsys@defobject{currentmarker}{\pgfqpoint{-0.048611in}{0.000000in}}{\pgfqpoint{-0.000000in}{0.000000in}}{%
\pgfpathmoveto{\pgfqpoint{-0.000000in}{0.000000in}}%
\pgfpathlineto{\pgfqpoint{-0.048611in}{0.000000in}}%
\pgfusepath{stroke,fill}%
}%
\begin{pgfscope}%
\pgfsys@transformshift{0.440443in}{3.414273in}%
\pgfsys@useobject{currentmarker}{}%
\end{pgfscope}%
\end{pgfscope}%
\begin{pgfscope}%
\definecolor{textcolor}{rgb}{0.000000,0.000000,0.000000}%
\pgfsetstrokecolor{textcolor}%
\pgfsetfillcolor{textcolor}%
\pgftext[x=0.284193in, y=3.372064in, left, base]{\color{textcolor}\rmfamily\fontsize{8.000000}{9.600000}\selectfont \(\displaystyle {0}\)}%
\end{pgfscope}%
\begin{pgfscope}%
\pgfsetbuttcap%
\pgfsetroundjoin%
\definecolor{currentfill}{rgb}{0.000000,0.000000,0.000000}%
\pgfsetfillcolor{currentfill}%
\pgfsetlinewidth{0.803000pt}%
\definecolor{currentstroke}{rgb}{0.000000,0.000000,0.000000}%
\pgfsetstrokecolor{currentstroke}%
\pgfsetdash{}{0pt}%
\pgfsys@defobject{currentmarker}{\pgfqpoint{-0.048611in}{0.000000in}}{\pgfqpoint{-0.000000in}{0.000000in}}{%
\pgfpathmoveto{\pgfqpoint{-0.000000in}{0.000000in}}%
\pgfpathlineto{\pgfqpoint{-0.048611in}{0.000000in}}%
\pgfusepath{stroke,fill}%
}%
\begin{pgfscope}%
\pgfsys@transformshift{0.440443in}{3.099020in}%
\pgfsys@useobject{currentmarker}{}%
\end{pgfscope}%
\end{pgfscope}%
\begin{pgfscope}%
\definecolor{textcolor}{rgb}{0.000000,0.000000,0.000000}%
\pgfsetstrokecolor{textcolor}%
\pgfsetfillcolor{textcolor}%
\pgftext[x=0.225164in, y=3.056811in, left, base]{\color{textcolor}\rmfamily\fontsize{8.000000}{9.600000}\selectfont \(\displaystyle {50}\)}%
\end{pgfscope}%
\begin{pgfscope}%
\pgfsetbuttcap%
\pgfsetroundjoin%
\definecolor{currentfill}{rgb}{0.000000,0.000000,0.000000}%
\pgfsetfillcolor{currentfill}%
\pgfsetlinewidth{0.803000pt}%
\definecolor{currentstroke}{rgb}{0.000000,0.000000,0.000000}%
\pgfsetstrokecolor{currentstroke}%
\pgfsetdash{}{0pt}%
\pgfsys@defobject{currentmarker}{\pgfqpoint{-0.048611in}{0.000000in}}{\pgfqpoint{-0.000000in}{0.000000in}}{%
\pgfpathmoveto{\pgfqpoint{-0.000000in}{0.000000in}}%
\pgfpathlineto{\pgfqpoint{-0.048611in}{0.000000in}}%
\pgfusepath{stroke,fill}%
}%
\begin{pgfscope}%
\pgfsys@transformshift{0.440443in}{2.783767in}%
\pgfsys@useobject{currentmarker}{}%
\end{pgfscope}%
\end{pgfscope}%
\begin{pgfscope}%
\definecolor{textcolor}{rgb}{0.000000,0.000000,0.000000}%
\pgfsetstrokecolor{textcolor}%
\pgfsetfillcolor{textcolor}%
\pgftext[x=0.166135in, y=2.741558in, left, base]{\color{textcolor}\rmfamily\fontsize{8.000000}{9.600000}\selectfont \(\displaystyle {100}\)}%
\end{pgfscope}%
\begin{pgfscope}%
\pgfsetbuttcap%
\pgfsetroundjoin%
\definecolor{currentfill}{rgb}{0.000000,0.000000,0.000000}%
\pgfsetfillcolor{currentfill}%
\pgfsetlinewidth{0.803000pt}%
\definecolor{currentstroke}{rgb}{0.000000,0.000000,0.000000}%
\pgfsetstrokecolor{currentstroke}%
\pgfsetdash{}{0pt}%
\pgfsys@defobject{currentmarker}{\pgfqpoint{-0.048611in}{0.000000in}}{\pgfqpoint{-0.000000in}{0.000000in}}{%
\pgfpathmoveto{\pgfqpoint{-0.000000in}{0.000000in}}%
\pgfpathlineto{\pgfqpoint{-0.048611in}{0.000000in}}%
\pgfusepath{stroke,fill}%
}%
\begin{pgfscope}%
\pgfsys@transformshift{0.440443in}{2.468514in}%
\pgfsys@useobject{currentmarker}{}%
\end{pgfscope}%
\end{pgfscope}%
\begin{pgfscope}%
\definecolor{textcolor}{rgb}{0.000000,0.000000,0.000000}%
\pgfsetstrokecolor{textcolor}%
\pgfsetfillcolor{textcolor}%
\pgftext[x=0.166135in, y=2.426305in, left, base]{\color{textcolor}\rmfamily\fontsize{8.000000}{9.600000}\selectfont \(\displaystyle {150}\)}%
\end{pgfscope}%
\begin{pgfscope}%
\definecolor{textcolor}{rgb}{0.000000,0.000000,0.000000}%
\pgfsetstrokecolor{textcolor}%
\pgfsetfillcolor{textcolor}%
\pgftext[x=0.110580in,y=2.862580in,,bottom,rotate=90.000000]{\color{textcolor}\rmfamily\fontsize{8.000000}{9.600000}\selectfont Elevation angle \(\displaystyle \theta\) (\(\displaystyle ^\circ\))}%
\end{pgfscope}%
\begin{pgfscope}%
\pgfsetrectcap%
\pgfsetmiterjoin%
\pgfsetlinewidth{0.803000pt}%
\definecolor{currentstroke}{rgb}{0.000000,0.000000,0.000000}%
\pgfsetstrokecolor{currentstroke}%
\pgfsetdash{}{0pt}%
\pgfpathmoveto{\pgfqpoint{0.440443in}{2.310888in}}%
\pgfpathlineto{\pgfqpoint{0.440443in}{3.414273in}}%
\pgfusepath{stroke}%
\end{pgfscope}%
\begin{pgfscope}%
\pgfsetrectcap%
\pgfsetmiterjoin%
\pgfsetlinewidth{0.803000pt}%
\definecolor{currentstroke}{rgb}{0.000000,0.000000,0.000000}%
\pgfsetstrokecolor{currentstroke}%
\pgfsetdash{}{0pt}%
\pgfpathmoveto{\pgfqpoint{2.678739in}{2.310888in}}%
\pgfpathlineto{\pgfqpoint{2.678739in}{3.414273in}}%
\pgfusepath{stroke}%
\end{pgfscope}%
\begin{pgfscope}%
\pgfsetrectcap%
\pgfsetmiterjoin%
\pgfsetlinewidth{0.803000pt}%
\definecolor{currentstroke}{rgb}{0.000000,0.000000,0.000000}%
\pgfsetstrokecolor{currentstroke}%
\pgfsetdash{}{0pt}%
\pgfpathmoveto{\pgfqpoint{0.440443in}{2.310888in}}%
\pgfpathlineto{\pgfqpoint{2.678739in}{2.310888in}}%
\pgfusepath{stroke}%
\end{pgfscope}%
\begin{pgfscope}%
\pgfsetrectcap%
\pgfsetmiterjoin%
\pgfsetlinewidth{0.803000pt}%
\definecolor{currentstroke}{rgb}{0.000000,0.000000,0.000000}%
\pgfsetstrokecolor{currentstroke}%
\pgfsetdash{}{0pt}%
\pgfpathmoveto{\pgfqpoint{0.440443in}{3.414273in}}%
\pgfpathlineto{\pgfqpoint{2.678739in}{3.414273in}}%
\pgfusepath{stroke}%
\end{pgfscope}%
\begin{pgfscope}%
\definecolor{textcolor}{rgb}{0.000000,0.000000,0.000000}%
\pgfsetstrokecolor{textcolor}%
\pgfsetfillcolor{textcolor}%
\pgftext[x=1.559591in,y=3.497606in,,base]{\color{textcolor}\rmfamily\fontsize{8.000000}{9.600000}\selectfont Vertical polarization}%
\end{pgfscope}%
\begin{pgfscope}%
\pgfsetbuttcap%
\pgfsetmiterjoin%
\definecolor{currentfill}{rgb}{1.000000,1.000000,1.000000}%
\pgfsetfillcolor{currentfill}%
\pgfsetlinewidth{0.000000pt}%
\definecolor{currentstroke}{rgb}{0.000000,0.000000,0.000000}%
\pgfsetstrokecolor{currentstroke}%
\pgfsetstrokeopacity{0.000000}%
\pgfsetdash{}{0pt}%
\pgfpathmoveto{\pgfqpoint{0.440443in}{0.370888in}}%
\pgfpathlineto{\pgfqpoint{2.678739in}{0.370888in}}%
\pgfpathlineto{\pgfqpoint{2.678739in}{1.474273in}}%
\pgfpathlineto{\pgfqpoint{0.440443in}{1.474273in}}%
\pgfpathlineto{\pgfqpoint{0.440443in}{0.370888in}}%
\pgfpathclose%
\pgfusepath{fill}%
\end{pgfscope}%
\begin{pgfscope}%
\pgfpathrectangle{\pgfqpoint{0.440443in}{0.370888in}}{\pgfqpoint{2.238295in}{1.103385in}}%
\pgfusepath{clip}%
\pgfsys@transformshift{0.440443in}{0.370888in}%
\pgftext[left,bottom]{\includegraphics[interpolate=true,width=2.240000in,height=1.110000in]{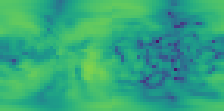}}%
\end{pgfscope}%
\begin{pgfscope}%
\pgfsetbuttcap%
\pgfsetroundjoin%
\definecolor{currentfill}{rgb}{0.000000,0.000000,0.000000}%
\pgfsetfillcolor{currentfill}%
\pgfsetlinewidth{0.803000pt}%
\definecolor{currentstroke}{rgb}{0.000000,0.000000,0.000000}%
\pgfsetstrokecolor{currentstroke}%
\pgfsetdash{}{0pt}%
\pgfsys@defobject{currentmarker}{\pgfqpoint{0.000000in}{-0.048611in}}{\pgfqpoint{0.000000in}{0.000000in}}{%
\pgfpathmoveto{\pgfqpoint{0.000000in}{0.000000in}}%
\pgfpathlineto{\pgfqpoint{0.000000in}{-0.048611in}}%
\pgfusepath{stroke,fill}%
}%
\begin{pgfscope}%
\pgfsys@transformshift{0.440443in}{0.370888in}%
\pgfsys@useobject{currentmarker}{}%
\end{pgfscope}%
\end{pgfscope}%
\begin{pgfscope}%
\definecolor{textcolor}{rgb}{0.000000,0.000000,0.000000}%
\pgfsetstrokecolor{textcolor}%
\pgfsetfillcolor{textcolor}%
\pgftext[x=0.440443in,y=0.273666in,,top]{\color{textcolor}\rmfamily\fontsize{8.000000}{9.600000}\selectfont \(\displaystyle {0}\)}%
\end{pgfscope}%
\begin{pgfscope}%
\pgfsetbuttcap%
\pgfsetroundjoin%
\definecolor{currentfill}{rgb}{0.000000,0.000000,0.000000}%
\pgfsetfillcolor{currentfill}%
\pgfsetlinewidth{0.803000pt}%
\definecolor{currentstroke}{rgb}{0.000000,0.000000,0.000000}%
\pgfsetstrokecolor{currentstroke}%
\pgfsetdash{}{0pt}%
\pgfsys@defobject{currentmarker}{\pgfqpoint{0.000000in}{-0.048611in}}{\pgfqpoint{0.000000in}{0.000000in}}{%
\pgfpathmoveto{\pgfqpoint{0.000000in}{0.000000in}}%
\pgfpathlineto{\pgfqpoint{0.000000in}{-0.048611in}}%
\pgfusepath{stroke,fill}%
}%
\begin{pgfscope}%
\pgfsys@transformshift{1.070949in}{0.370888in}%
\pgfsys@useobject{currentmarker}{}%
\end{pgfscope}%
\end{pgfscope}%
\begin{pgfscope}%
\definecolor{textcolor}{rgb}{0.000000,0.000000,0.000000}%
\pgfsetstrokecolor{textcolor}%
\pgfsetfillcolor{textcolor}%
\pgftext[x=1.070949in,y=0.273666in,,top]{\color{textcolor}\rmfamily\fontsize{8.000000}{9.600000}\selectfont \(\displaystyle {100}\)}%
\end{pgfscope}%
\begin{pgfscope}%
\pgfsetbuttcap%
\pgfsetroundjoin%
\definecolor{currentfill}{rgb}{0.000000,0.000000,0.000000}%
\pgfsetfillcolor{currentfill}%
\pgfsetlinewidth{0.803000pt}%
\definecolor{currentstroke}{rgb}{0.000000,0.000000,0.000000}%
\pgfsetstrokecolor{currentstroke}%
\pgfsetdash{}{0pt}%
\pgfsys@defobject{currentmarker}{\pgfqpoint{0.000000in}{-0.048611in}}{\pgfqpoint{0.000000in}{0.000000in}}{%
\pgfpathmoveto{\pgfqpoint{0.000000in}{0.000000in}}%
\pgfpathlineto{\pgfqpoint{0.000000in}{-0.048611in}}%
\pgfusepath{stroke,fill}%
}%
\begin{pgfscope}%
\pgfsys@transformshift{1.701455in}{0.370888in}%
\pgfsys@useobject{currentmarker}{}%
\end{pgfscope}%
\end{pgfscope}%
\begin{pgfscope}%
\definecolor{textcolor}{rgb}{0.000000,0.000000,0.000000}%
\pgfsetstrokecolor{textcolor}%
\pgfsetfillcolor{textcolor}%
\pgftext[x=1.701455in,y=0.273666in,,top]{\color{textcolor}\rmfamily\fontsize{8.000000}{9.600000}\selectfont \(\displaystyle {200}\)}%
\end{pgfscope}%
\begin{pgfscope}%
\pgfsetbuttcap%
\pgfsetroundjoin%
\definecolor{currentfill}{rgb}{0.000000,0.000000,0.000000}%
\pgfsetfillcolor{currentfill}%
\pgfsetlinewidth{0.803000pt}%
\definecolor{currentstroke}{rgb}{0.000000,0.000000,0.000000}%
\pgfsetstrokecolor{currentstroke}%
\pgfsetdash{}{0pt}%
\pgfsys@defobject{currentmarker}{\pgfqpoint{0.000000in}{-0.048611in}}{\pgfqpoint{0.000000in}{0.000000in}}{%
\pgfpathmoveto{\pgfqpoint{0.000000in}{0.000000in}}%
\pgfpathlineto{\pgfqpoint{0.000000in}{-0.048611in}}%
\pgfusepath{stroke,fill}%
}%
\begin{pgfscope}%
\pgfsys@transformshift{2.331960in}{0.370888in}%
\pgfsys@useobject{currentmarker}{}%
\end{pgfscope}%
\end{pgfscope}%
\begin{pgfscope}%
\definecolor{textcolor}{rgb}{0.000000,0.000000,0.000000}%
\pgfsetstrokecolor{textcolor}%
\pgfsetfillcolor{textcolor}%
\pgftext[x=2.331960in,y=0.273666in,,top]{\color{textcolor}\rmfamily\fontsize{8.000000}{9.600000}\selectfont \(\displaystyle {300}\)}%
\end{pgfscope}%
\begin{pgfscope}%
\definecolor{textcolor}{rgb}{0.000000,0.000000,0.000000}%
\pgfsetstrokecolor{textcolor}%
\pgfsetfillcolor{textcolor}%
\pgftext[x=1.559591in,y=0.110580in,,top]{\color{textcolor}\rmfamily\fontsize{8.000000}{9.600000}\selectfont Azimuth angle \(\displaystyle \phi\) (\(\displaystyle ^\circ\))}%
\end{pgfscope}%
\begin{pgfscope}%
\pgfsetbuttcap%
\pgfsetroundjoin%
\definecolor{currentfill}{rgb}{0.000000,0.000000,0.000000}%
\pgfsetfillcolor{currentfill}%
\pgfsetlinewidth{0.803000pt}%
\definecolor{currentstroke}{rgb}{0.000000,0.000000,0.000000}%
\pgfsetstrokecolor{currentstroke}%
\pgfsetdash{}{0pt}%
\pgfsys@defobject{currentmarker}{\pgfqpoint{-0.048611in}{0.000000in}}{\pgfqpoint{-0.000000in}{0.000000in}}{%
\pgfpathmoveto{\pgfqpoint{-0.000000in}{0.000000in}}%
\pgfpathlineto{\pgfqpoint{-0.048611in}{0.000000in}}%
\pgfusepath{stroke,fill}%
}%
\begin{pgfscope}%
\pgfsys@transformshift{0.440443in}{1.474273in}%
\pgfsys@useobject{currentmarker}{}%
\end{pgfscope}%
\end{pgfscope}%
\begin{pgfscope}%
\definecolor{textcolor}{rgb}{0.000000,0.000000,0.000000}%
\pgfsetstrokecolor{textcolor}%
\pgfsetfillcolor{textcolor}%
\pgftext[x=0.284193in, y=1.432064in, left, base]{\color{textcolor}\rmfamily\fontsize{8.000000}{9.600000}\selectfont \(\displaystyle {0}\)}%
\end{pgfscope}%
\begin{pgfscope}%
\pgfsetbuttcap%
\pgfsetroundjoin%
\definecolor{currentfill}{rgb}{0.000000,0.000000,0.000000}%
\pgfsetfillcolor{currentfill}%
\pgfsetlinewidth{0.803000pt}%
\definecolor{currentstroke}{rgb}{0.000000,0.000000,0.000000}%
\pgfsetstrokecolor{currentstroke}%
\pgfsetdash{}{0pt}%
\pgfsys@defobject{currentmarker}{\pgfqpoint{-0.048611in}{0.000000in}}{\pgfqpoint{-0.000000in}{0.000000in}}{%
\pgfpathmoveto{\pgfqpoint{-0.000000in}{0.000000in}}%
\pgfpathlineto{\pgfqpoint{-0.048611in}{0.000000in}}%
\pgfusepath{stroke,fill}%
}%
\begin{pgfscope}%
\pgfsys@transformshift{0.440443in}{1.159020in}%
\pgfsys@useobject{currentmarker}{}%
\end{pgfscope}%
\end{pgfscope}%
\begin{pgfscope}%
\definecolor{textcolor}{rgb}{0.000000,0.000000,0.000000}%
\pgfsetstrokecolor{textcolor}%
\pgfsetfillcolor{textcolor}%
\pgftext[x=0.225164in, y=1.116811in, left, base]{\color{textcolor}\rmfamily\fontsize{8.000000}{9.600000}\selectfont \(\displaystyle {50}\)}%
\end{pgfscope}%
\begin{pgfscope}%
\pgfsetbuttcap%
\pgfsetroundjoin%
\definecolor{currentfill}{rgb}{0.000000,0.000000,0.000000}%
\pgfsetfillcolor{currentfill}%
\pgfsetlinewidth{0.803000pt}%
\definecolor{currentstroke}{rgb}{0.000000,0.000000,0.000000}%
\pgfsetstrokecolor{currentstroke}%
\pgfsetdash{}{0pt}%
\pgfsys@defobject{currentmarker}{\pgfqpoint{-0.048611in}{0.000000in}}{\pgfqpoint{-0.000000in}{0.000000in}}{%
\pgfpathmoveto{\pgfqpoint{-0.000000in}{0.000000in}}%
\pgfpathlineto{\pgfqpoint{-0.048611in}{0.000000in}}%
\pgfusepath{stroke,fill}%
}%
\begin{pgfscope}%
\pgfsys@transformshift{0.440443in}{0.843767in}%
\pgfsys@useobject{currentmarker}{}%
\end{pgfscope}%
\end{pgfscope}%
\begin{pgfscope}%
\definecolor{textcolor}{rgb}{0.000000,0.000000,0.000000}%
\pgfsetstrokecolor{textcolor}%
\pgfsetfillcolor{textcolor}%
\pgftext[x=0.166135in, y=0.801558in, left, base]{\color{textcolor}\rmfamily\fontsize{8.000000}{9.600000}\selectfont \(\displaystyle {100}\)}%
\end{pgfscope}%
\begin{pgfscope}%
\pgfsetbuttcap%
\pgfsetroundjoin%
\definecolor{currentfill}{rgb}{0.000000,0.000000,0.000000}%
\pgfsetfillcolor{currentfill}%
\pgfsetlinewidth{0.803000pt}%
\definecolor{currentstroke}{rgb}{0.000000,0.000000,0.000000}%
\pgfsetstrokecolor{currentstroke}%
\pgfsetdash{}{0pt}%
\pgfsys@defobject{currentmarker}{\pgfqpoint{-0.048611in}{0.000000in}}{\pgfqpoint{-0.000000in}{0.000000in}}{%
\pgfpathmoveto{\pgfqpoint{-0.000000in}{0.000000in}}%
\pgfpathlineto{\pgfqpoint{-0.048611in}{0.000000in}}%
\pgfusepath{stroke,fill}%
}%
\begin{pgfscope}%
\pgfsys@transformshift{0.440443in}{0.528514in}%
\pgfsys@useobject{currentmarker}{}%
\end{pgfscope}%
\end{pgfscope}%
\begin{pgfscope}%
\definecolor{textcolor}{rgb}{0.000000,0.000000,0.000000}%
\pgfsetstrokecolor{textcolor}%
\pgfsetfillcolor{textcolor}%
\pgftext[x=0.166135in, y=0.486305in, left, base]{\color{textcolor}\rmfamily\fontsize{8.000000}{9.600000}\selectfont \(\displaystyle {150}\)}%
\end{pgfscope}%
\begin{pgfscope}%
\definecolor{textcolor}{rgb}{0.000000,0.000000,0.000000}%
\pgfsetstrokecolor{textcolor}%
\pgfsetfillcolor{textcolor}%
\pgftext[x=0.110580in,y=0.922580in,,bottom,rotate=90.000000]{\color{textcolor}\rmfamily\fontsize{8.000000}{9.600000}\selectfont Elevation angle \(\displaystyle \theta\) (\(\displaystyle ^\circ\))}%
\end{pgfscope}%
\begin{pgfscope}%
\pgfsetrectcap%
\pgfsetmiterjoin%
\pgfsetlinewidth{0.803000pt}%
\definecolor{currentstroke}{rgb}{0.000000,0.000000,0.000000}%
\pgfsetstrokecolor{currentstroke}%
\pgfsetdash{}{0pt}%
\pgfpathmoveto{\pgfqpoint{0.440443in}{0.370888in}}%
\pgfpathlineto{\pgfqpoint{0.440443in}{1.474273in}}%
\pgfusepath{stroke}%
\end{pgfscope}%
\begin{pgfscope}%
\pgfsetrectcap%
\pgfsetmiterjoin%
\pgfsetlinewidth{0.803000pt}%
\definecolor{currentstroke}{rgb}{0.000000,0.000000,0.000000}%
\pgfsetstrokecolor{currentstroke}%
\pgfsetdash{}{0pt}%
\pgfpathmoveto{\pgfqpoint{2.678739in}{0.370888in}}%
\pgfpathlineto{\pgfqpoint{2.678739in}{1.474273in}}%
\pgfusepath{stroke}%
\end{pgfscope}%
\begin{pgfscope}%
\pgfsetrectcap%
\pgfsetmiterjoin%
\pgfsetlinewidth{0.803000pt}%
\definecolor{currentstroke}{rgb}{0.000000,0.000000,0.000000}%
\pgfsetstrokecolor{currentstroke}%
\pgfsetdash{}{0pt}%
\pgfpathmoveto{\pgfqpoint{0.440443in}{0.370888in}}%
\pgfpathlineto{\pgfqpoint{2.678739in}{0.370888in}}%
\pgfusepath{stroke}%
\end{pgfscope}%
\begin{pgfscope}%
\pgfsetrectcap%
\pgfsetmiterjoin%
\pgfsetlinewidth{0.803000pt}%
\definecolor{currentstroke}{rgb}{0.000000,0.000000,0.000000}%
\pgfsetstrokecolor{currentstroke}%
\pgfsetdash{}{0pt}%
\pgfpathmoveto{\pgfqpoint{0.440443in}{1.474273in}}%
\pgfpathlineto{\pgfqpoint{2.678739in}{1.474273in}}%
\pgfusepath{stroke}%
\end{pgfscope}%
\begin{pgfscope}%
\definecolor{textcolor}{rgb}{0.000000,0.000000,0.000000}%
\pgfsetstrokecolor{textcolor}%
\pgfsetfillcolor{textcolor}%
\pgftext[x=1.559591in,y=1.557606in,,base]{\color{textcolor}\rmfamily\fontsize{8.000000}{9.600000}\selectfont Horizontal polarization}%
\end{pgfscope}%
\begin{pgfscope}%
\pgfsetbuttcap%
\pgfsetmiterjoin%
\definecolor{currentfill}{rgb}{1.000000,1.000000,1.000000}%
\pgfsetfillcolor{currentfill}%
\pgfsetlinewidth{0.000000pt}%
\definecolor{currentstroke}{rgb}{0.000000,0.000000,0.000000}%
\pgfsetstrokecolor{currentstroke}%
\pgfsetstrokeopacity{0.000000}%
\pgfsetdash{}{0pt}%
\pgfpathmoveto{\pgfqpoint{2.818632in}{2.072041in}}%
\pgfpathlineto{\pgfqpoint{2.897686in}{2.072041in}}%
\pgfpathlineto{\pgfqpoint{2.897686in}{3.653120in}}%
\pgfpathlineto{\pgfqpoint{2.818632in}{3.653120in}}%
\pgfpathlineto{\pgfqpoint{2.818632in}{2.072041in}}%
\pgfpathclose%
\pgfusepath{fill}%
\end{pgfscope}%
\begin{pgfscope}%
\pgfsys@transformshift{2.820000in}{2.073120in}%
\pgftext[left,bottom]{\includegraphics[interpolate=true,width=0.080000in,height=1.580000in]{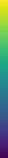}}%
\end{pgfscope}%
\begin{pgfscope}%
\pgfsetbuttcap%
\pgfsetroundjoin%
\definecolor{currentfill}{rgb}{0.000000,0.000000,0.000000}%
\pgfsetfillcolor{currentfill}%
\pgfsetlinewidth{0.803000pt}%
\definecolor{currentstroke}{rgb}{0.000000,0.000000,0.000000}%
\pgfsetstrokecolor{currentstroke}%
\pgfsetdash{}{0pt}%
\pgfsys@defobject{currentmarker}{\pgfqpoint{0.000000in}{0.000000in}}{\pgfqpoint{0.048611in}{0.000000in}}{%
\pgfpathmoveto{\pgfqpoint{0.000000in}{0.000000in}}%
\pgfpathlineto{\pgfqpoint{0.048611in}{0.000000in}}%
\pgfusepath{stroke,fill}%
}%
\begin{pgfscope}%
\pgfsys@transformshift{2.897686in}{2.256814in}%
\pgfsys@useobject{currentmarker}{}%
\end{pgfscope}%
\end{pgfscope}%
\begin{pgfscope}%
\definecolor{textcolor}{rgb}{0.000000,0.000000,0.000000}%
\pgfsetstrokecolor{textcolor}%
\pgfsetfillcolor{textcolor}%
\pgftext[x=2.994908in, y=2.214604in, left, base]{\color{textcolor}\rmfamily\fontsize{8.000000}{9.600000}\selectfont \(\displaystyle {\ensuremath{-}50}\)}%
\end{pgfscope}%
\begin{pgfscope}%
\pgfsetbuttcap%
\pgfsetroundjoin%
\definecolor{currentfill}{rgb}{0.000000,0.000000,0.000000}%
\pgfsetfillcolor{currentfill}%
\pgfsetlinewidth{0.803000pt}%
\definecolor{currentstroke}{rgb}{0.000000,0.000000,0.000000}%
\pgfsetstrokecolor{currentstroke}%
\pgfsetdash{}{0pt}%
\pgfsys@defobject{currentmarker}{\pgfqpoint{0.000000in}{0.000000in}}{\pgfqpoint{0.048611in}{0.000000in}}{%
\pgfpathmoveto{\pgfqpoint{0.000000in}{0.000000in}}%
\pgfpathlineto{\pgfqpoint{0.048611in}{0.000000in}}%
\pgfusepath{stroke,fill}%
}%
\begin{pgfscope}%
\pgfsys@transformshift{2.897686in}{2.538121in}%
\pgfsys@useobject{currentmarker}{}%
\end{pgfscope}%
\end{pgfscope}%
\begin{pgfscope}%
\definecolor{textcolor}{rgb}{0.000000,0.000000,0.000000}%
\pgfsetstrokecolor{textcolor}%
\pgfsetfillcolor{textcolor}%
\pgftext[x=2.994908in, y=2.495911in, left, base]{\color{textcolor}\rmfamily\fontsize{8.000000}{9.600000}\selectfont \(\displaystyle {\ensuremath{-}40}\)}%
\end{pgfscope}%
\begin{pgfscope}%
\pgfsetbuttcap%
\pgfsetroundjoin%
\definecolor{currentfill}{rgb}{0.000000,0.000000,0.000000}%
\pgfsetfillcolor{currentfill}%
\pgfsetlinewidth{0.803000pt}%
\definecolor{currentstroke}{rgb}{0.000000,0.000000,0.000000}%
\pgfsetstrokecolor{currentstroke}%
\pgfsetdash{}{0pt}%
\pgfsys@defobject{currentmarker}{\pgfqpoint{0.000000in}{0.000000in}}{\pgfqpoint{0.048611in}{0.000000in}}{%
\pgfpathmoveto{\pgfqpoint{0.000000in}{0.000000in}}%
\pgfpathlineto{\pgfqpoint{0.048611in}{0.000000in}}%
\pgfusepath{stroke,fill}%
}%
\begin{pgfscope}%
\pgfsys@transformshift{2.897686in}{2.819428in}%
\pgfsys@useobject{currentmarker}{}%
\end{pgfscope}%
\end{pgfscope}%
\begin{pgfscope}%
\definecolor{textcolor}{rgb}{0.000000,0.000000,0.000000}%
\pgfsetstrokecolor{textcolor}%
\pgfsetfillcolor{textcolor}%
\pgftext[x=2.994908in, y=2.777218in, left, base]{\color{textcolor}\rmfamily\fontsize{8.000000}{9.600000}\selectfont \(\displaystyle {\ensuremath{-}30}\)}%
\end{pgfscope}%
\begin{pgfscope}%
\pgfsetbuttcap%
\pgfsetroundjoin%
\definecolor{currentfill}{rgb}{0.000000,0.000000,0.000000}%
\pgfsetfillcolor{currentfill}%
\pgfsetlinewidth{0.803000pt}%
\definecolor{currentstroke}{rgb}{0.000000,0.000000,0.000000}%
\pgfsetstrokecolor{currentstroke}%
\pgfsetdash{}{0pt}%
\pgfsys@defobject{currentmarker}{\pgfqpoint{0.000000in}{0.000000in}}{\pgfqpoint{0.048611in}{0.000000in}}{%
\pgfpathmoveto{\pgfqpoint{0.000000in}{0.000000in}}%
\pgfpathlineto{\pgfqpoint{0.048611in}{0.000000in}}%
\pgfusepath{stroke,fill}%
}%
\begin{pgfscope}%
\pgfsys@transformshift{2.897686in}{3.100735in}%
\pgfsys@useobject{currentmarker}{}%
\end{pgfscope}%
\end{pgfscope}%
\begin{pgfscope}%
\definecolor{textcolor}{rgb}{0.000000,0.000000,0.000000}%
\pgfsetstrokecolor{textcolor}%
\pgfsetfillcolor{textcolor}%
\pgftext[x=2.994908in, y=3.058525in, left, base]{\color{textcolor}\rmfamily\fontsize{8.000000}{9.600000}\selectfont \(\displaystyle {\ensuremath{-}20}\)}%
\end{pgfscope}%
\begin{pgfscope}%
\pgfsetbuttcap%
\pgfsetroundjoin%
\definecolor{currentfill}{rgb}{0.000000,0.000000,0.000000}%
\pgfsetfillcolor{currentfill}%
\pgfsetlinewidth{0.803000pt}%
\definecolor{currentstroke}{rgb}{0.000000,0.000000,0.000000}%
\pgfsetstrokecolor{currentstroke}%
\pgfsetdash{}{0pt}%
\pgfsys@defobject{currentmarker}{\pgfqpoint{0.000000in}{0.000000in}}{\pgfqpoint{0.048611in}{0.000000in}}{%
\pgfpathmoveto{\pgfqpoint{0.000000in}{0.000000in}}%
\pgfpathlineto{\pgfqpoint{0.048611in}{0.000000in}}%
\pgfusepath{stroke,fill}%
}%
\begin{pgfscope}%
\pgfsys@transformshift{2.897686in}{3.382042in}%
\pgfsys@useobject{currentmarker}{}%
\end{pgfscope}%
\end{pgfscope}%
\begin{pgfscope}%
\definecolor{textcolor}{rgb}{0.000000,0.000000,0.000000}%
\pgfsetstrokecolor{textcolor}%
\pgfsetfillcolor{textcolor}%
\pgftext[x=2.994908in, y=3.339833in, left, base]{\color{textcolor}\rmfamily\fontsize{8.000000}{9.600000}\selectfont \(\displaystyle {\ensuremath{-}10}\)}%
\end{pgfscope}%
\begin{pgfscope}%
\definecolor{textcolor}{rgb}{0.000000,0.000000,0.000000}%
\pgfsetstrokecolor{textcolor}%
\pgfsetfillcolor{textcolor}%
\pgftext[x=3.260343in,y=2.862580in,,top,rotate=90.000000]{\color{textcolor}\rmfamily\fontsize{8.000000}{9.600000}\selectfont Gain (dB)}%
\end{pgfscope}%
\begin{pgfscope}%
\pgfsetrectcap%
\pgfsetmiterjoin%
\pgfsetlinewidth{0.803000pt}%
\definecolor{currentstroke}{rgb}{0.000000,0.000000,0.000000}%
\pgfsetstrokecolor{currentstroke}%
\pgfsetdash{}{0pt}%
\pgfpathmoveto{\pgfqpoint{2.818632in}{2.072041in}}%
\pgfpathlineto{\pgfqpoint{2.858159in}{2.072041in}}%
\pgfpathlineto{\pgfqpoint{2.897686in}{2.072041in}}%
\pgfpathlineto{\pgfqpoint{2.897686in}{3.653120in}}%
\pgfpathlineto{\pgfqpoint{2.858159in}{3.653120in}}%
\pgfpathlineto{\pgfqpoint{2.818632in}{3.653120in}}%
\pgfpathlineto{\pgfqpoint{2.818632in}{2.072041in}}%
\pgfpathclose%
\pgfusepath{stroke}%
\end{pgfscope}%
\begin{pgfscope}%
\pgfsetbuttcap%
\pgfsetmiterjoin%
\definecolor{currentfill}{rgb}{1.000000,1.000000,1.000000}%
\pgfsetfillcolor{currentfill}%
\pgfsetlinewidth{0.000000pt}%
\definecolor{currentstroke}{rgb}{0.000000,0.000000,0.000000}%
\pgfsetstrokecolor{currentstroke}%
\pgfsetstrokeopacity{0.000000}%
\pgfsetdash{}{0pt}%
\pgfpathmoveto{\pgfqpoint{2.818632in}{0.132041in}}%
\pgfpathlineto{\pgfqpoint{2.897686in}{0.132041in}}%
\pgfpathlineto{\pgfqpoint{2.897686in}{1.713120in}}%
\pgfpathlineto{\pgfqpoint{2.818632in}{1.713120in}}%
\pgfpathlineto{\pgfqpoint{2.818632in}{0.132041in}}%
\pgfpathclose%
\pgfusepath{fill}%
\end{pgfscope}%
\begin{pgfscope}%
\pgfsys@transformshift{2.820000in}{0.133120in}%
\pgftext[left,bottom]{\includegraphics[interpolate=true,width=0.080000in,height=1.580000in]{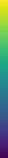}}%
\end{pgfscope}%
\begin{pgfscope}%
\pgfsetbuttcap%
\pgfsetroundjoin%
\definecolor{currentfill}{rgb}{0.000000,0.000000,0.000000}%
\pgfsetfillcolor{currentfill}%
\pgfsetlinewidth{0.803000pt}%
\definecolor{currentstroke}{rgb}{0.000000,0.000000,0.000000}%
\pgfsetstrokecolor{currentstroke}%
\pgfsetdash{}{0pt}%
\pgfsys@defobject{currentmarker}{\pgfqpoint{0.000000in}{0.000000in}}{\pgfqpoint{0.048611in}{0.000000in}}{%
\pgfpathmoveto{\pgfqpoint{0.000000in}{0.000000in}}%
\pgfpathlineto{\pgfqpoint{0.048611in}{0.000000in}}%
\pgfusepath{stroke,fill}%
}%
\begin{pgfscope}%
\pgfsys@transformshift{2.897686in}{0.316814in}%
\pgfsys@useobject{currentmarker}{}%
\end{pgfscope}%
\end{pgfscope}%
\begin{pgfscope}%
\definecolor{textcolor}{rgb}{0.000000,0.000000,0.000000}%
\pgfsetstrokecolor{textcolor}%
\pgfsetfillcolor{textcolor}%
\pgftext[x=2.994908in, y=0.274604in, left, base]{\color{textcolor}\rmfamily\fontsize{8.000000}{9.600000}\selectfont \(\displaystyle {\ensuremath{-}50}\)}%
\end{pgfscope}%
\begin{pgfscope}%
\pgfsetbuttcap%
\pgfsetroundjoin%
\definecolor{currentfill}{rgb}{0.000000,0.000000,0.000000}%
\pgfsetfillcolor{currentfill}%
\pgfsetlinewidth{0.803000pt}%
\definecolor{currentstroke}{rgb}{0.000000,0.000000,0.000000}%
\pgfsetstrokecolor{currentstroke}%
\pgfsetdash{}{0pt}%
\pgfsys@defobject{currentmarker}{\pgfqpoint{0.000000in}{0.000000in}}{\pgfqpoint{0.048611in}{0.000000in}}{%
\pgfpathmoveto{\pgfqpoint{0.000000in}{0.000000in}}%
\pgfpathlineto{\pgfqpoint{0.048611in}{0.000000in}}%
\pgfusepath{stroke,fill}%
}%
\begin{pgfscope}%
\pgfsys@transformshift{2.897686in}{0.598121in}%
\pgfsys@useobject{currentmarker}{}%
\end{pgfscope}%
\end{pgfscope}%
\begin{pgfscope}%
\definecolor{textcolor}{rgb}{0.000000,0.000000,0.000000}%
\pgfsetstrokecolor{textcolor}%
\pgfsetfillcolor{textcolor}%
\pgftext[x=2.994908in, y=0.555911in, left, base]{\color{textcolor}\rmfamily\fontsize{8.000000}{9.600000}\selectfont \(\displaystyle {\ensuremath{-}40}\)}%
\end{pgfscope}%
\begin{pgfscope}%
\pgfsetbuttcap%
\pgfsetroundjoin%
\definecolor{currentfill}{rgb}{0.000000,0.000000,0.000000}%
\pgfsetfillcolor{currentfill}%
\pgfsetlinewidth{0.803000pt}%
\definecolor{currentstroke}{rgb}{0.000000,0.000000,0.000000}%
\pgfsetstrokecolor{currentstroke}%
\pgfsetdash{}{0pt}%
\pgfsys@defobject{currentmarker}{\pgfqpoint{0.000000in}{0.000000in}}{\pgfqpoint{0.048611in}{0.000000in}}{%
\pgfpathmoveto{\pgfqpoint{0.000000in}{0.000000in}}%
\pgfpathlineto{\pgfqpoint{0.048611in}{0.000000in}}%
\pgfusepath{stroke,fill}%
}%
\begin{pgfscope}%
\pgfsys@transformshift{2.897686in}{0.879428in}%
\pgfsys@useobject{currentmarker}{}%
\end{pgfscope}%
\end{pgfscope}%
\begin{pgfscope}%
\definecolor{textcolor}{rgb}{0.000000,0.000000,0.000000}%
\pgfsetstrokecolor{textcolor}%
\pgfsetfillcolor{textcolor}%
\pgftext[x=2.994908in, y=0.837218in, left, base]{\color{textcolor}\rmfamily\fontsize{8.000000}{9.600000}\selectfont \(\displaystyle {\ensuremath{-}30}\)}%
\end{pgfscope}%
\begin{pgfscope}%
\pgfsetbuttcap%
\pgfsetroundjoin%
\definecolor{currentfill}{rgb}{0.000000,0.000000,0.000000}%
\pgfsetfillcolor{currentfill}%
\pgfsetlinewidth{0.803000pt}%
\definecolor{currentstroke}{rgb}{0.000000,0.000000,0.000000}%
\pgfsetstrokecolor{currentstroke}%
\pgfsetdash{}{0pt}%
\pgfsys@defobject{currentmarker}{\pgfqpoint{0.000000in}{0.000000in}}{\pgfqpoint{0.048611in}{0.000000in}}{%
\pgfpathmoveto{\pgfqpoint{0.000000in}{0.000000in}}%
\pgfpathlineto{\pgfqpoint{0.048611in}{0.000000in}}%
\pgfusepath{stroke,fill}%
}%
\begin{pgfscope}%
\pgfsys@transformshift{2.897686in}{1.160735in}%
\pgfsys@useobject{currentmarker}{}%
\end{pgfscope}%
\end{pgfscope}%
\begin{pgfscope}%
\definecolor{textcolor}{rgb}{0.000000,0.000000,0.000000}%
\pgfsetstrokecolor{textcolor}%
\pgfsetfillcolor{textcolor}%
\pgftext[x=2.994908in, y=1.118525in, left, base]{\color{textcolor}\rmfamily\fontsize{8.000000}{9.600000}\selectfont \(\displaystyle {\ensuremath{-}20}\)}%
\end{pgfscope}%
\begin{pgfscope}%
\pgfsetbuttcap%
\pgfsetroundjoin%
\definecolor{currentfill}{rgb}{0.000000,0.000000,0.000000}%
\pgfsetfillcolor{currentfill}%
\pgfsetlinewidth{0.803000pt}%
\definecolor{currentstroke}{rgb}{0.000000,0.000000,0.000000}%
\pgfsetstrokecolor{currentstroke}%
\pgfsetdash{}{0pt}%
\pgfsys@defobject{currentmarker}{\pgfqpoint{0.000000in}{0.000000in}}{\pgfqpoint{0.048611in}{0.000000in}}{%
\pgfpathmoveto{\pgfqpoint{0.000000in}{0.000000in}}%
\pgfpathlineto{\pgfqpoint{0.048611in}{0.000000in}}%
\pgfusepath{stroke,fill}%
}%
\begin{pgfscope}%
\pgfsys@transformshift{2.897686in}{1.442042in}%
\pgfsys@useobject{currentmarker}{}%
\end{pgfscope}%
\end{pgfscope}%
\begin{pgfscope}%
\definecolor{textcolor}{rgb}{0.000000,0.000000,0.000000}%
\pgfsetstrokecolor{textcolor}%
\pgfsetfillcolor{textcolor}%
\pgftext[x=2.994908in, y=1.399833in, left, base]{\color{textcolor}\rmfamily\fontsize{8.000000}{9.600000}\selectfont \(\displaystyle {\ensuremath{-}10}\)}%
\end{pgfscope}%
\begin{pgfscope}%
\definecolor{textcolor}{rgb}{0.000000,0.000000,0.000000}%
\pgfsetstrokecolor{textcolor}%
\pgfsetfillcolor{textcolor}%
\pgftext[x=3.260343in,y=0.922580in,,top,rotate=90.000000]{\color{textcolor}\rmfamily\fontsize{8.000000}{9.600000}\selectfont Gain (dB)}%
\end{pgfscope}%
\begin{pgfscope}%
\pgfsetrectcap%
\pgfsetmiterjoin%
\pgfsetlinewidth{0.803000pt}%
\definecolor{currentstroke}{rgb}{0.000000,0.000000,0.000000}%
\pgfsetstrokecolor{currentstroke}%
\pgfsetdash{}{0pt}%
\pgfpathmoveto{\pgfqpoint{2.818632in}{0.132041in}}%
\pgfpathlineto{\pgfqpoint{2.858159in}{0.132041in}}%
\pgfpathlineto{\pgfqpoint{2.897686in}{0.132041in}}%
\pgfpathlineto{\pgfqpoint{2.897686in}{1.713120in}}%
\pgfpathlineto{\pgfqpoint{2.858159in}{1.713120in}}%
\pgfpathlineto{\pgfqpoint{2.818632in}{1.713120in}}%
\pgfpathlineto{\pgfqpoint{2.818632in}{0.132041in}}%
\pgfpathclose%
\pgfusepath{stroke}%
\end{pgfscope}%
\end{pgfpicture}%
\makeatother%
\endgroup%

%% file: scenarios2.tex
%
%
\definecolor{mycolor1}{rgb}{0.30100,0.74500,0.93300}%
\begin{tikzpicture}

\begin{axis}[%
width=2.75in,
height=2.75in,
scale only axis,
xmin=-0.3,
xmax=6.5,
ymin=-0.3,
ymax=6.5,
xlabel={x (m)},
ylabel={y (m)},
axis background/.style={fill=white},
legend style={at={(0.5,1.01)}, anchor=south, legend cell align=left, align=left, draw=white!15!black, column sep=0.1cm},
legend columns=2,
transpose legend
]

\draw[pattern=north west lines, pattern color=brown!50] (-0.3,-0.3) rectangle (6.5, 6.5);

\addplot [color=black, forget plot, fill=white]
  table[row sep=crcr]{%
0	6.4\\
0	0\\
6.03	0\\
6.03	6.4\\
} -- cycle;
\addplot [color=black, line width=2.0pt, forget plot]
  table[row sep=crcr]{%
0.461	6.131\\
0.245	5.964\\
};
\addplot [color=black, line width=2.0pt, forget plot]
  table[row sep=crcr]{%
4.37	6.147\\
4.593	6.016\\
};
\addplot [color=black, line width=2.0pt, forget plot]
  table[row sep=crcr]{%
0.255	0.509\\
0.456	0.361\\
};
\addplot [color=black, line width=2.0pt, forget plot]
  table[row sep=crcr]{%
4.2908	0.192\\
4.4888	0.3311\\
};

\addplot[only marks, mark=+, mark options={}, mark size=1.5000pt, draw=mycolor1] table[row sep=crcr]{%
x	y\\
2.79158204835889	3.42066380559515\\
2.43330399057316	3.36825590567349\\
0.659833990925668	3.46613231651656\\
0.716431074207686	3.11691274903237\\
3.61528450084536	3.49728021364102\\
2.33193630206724	3.08064391298693\\
2.51608287460544	3.17584478427398\\
1.1176759866216	3.39163927637717\\
1.97696609917887	3.32763725310126\\
3.73366311311717	3.09399134836676\\
3.55759016307946	3.06457767099456\\
2.10075382025712	3.0954922099107\\
2.69009822405675	3.33334817515974\\
0.932810962462944	2.99920214235528\\
1.30155214077799	3.39581242034536\\
3.59685992393655	3.12572365333124\\
2.51126644377879	3.34412776380951\\
0.93923676020857	3.26417497584496\\
0.704944574914667	3.27509556071309\\
2.74535216139814	3.47361597390361\\
2.83104193146693	3.32775176720999\\
2.70131229771889	3.28431055669898\\
1.99047012311222	3.070194668036\\
2.77190763440487	3.35173697337568\\
3.58032419653468	3.3611487747725\\
2.51101390598942	3.29423271542702\\
2.14864308848298	3.4784826293794\\
2.52703181162201	3.37592484146936\\
2.48950676611896	3.22562699965011\\
2.57986126293684	3.29256785804293\\
1.47570805075003	3.38886087040311\\
2.66615871855787	3.1412092754864\\
2.8096652646202	3.3100689508031\\
1.76550008488757	2.5356382567965\\
2.57198603466691	3.30803158672641\\
2.65667937614725	3.1688439628536\\
2.47844367470351	3.21433883087484\\
3.41536298168846	3.30027767928506\\
2.22292381398992	3.07218618786644\\
0.470276504651846	3.49443101554043\\
3.64650600360052	3.51803686314234\\
0.407040584553394	3.09726315327698\\
0.599935889601835	3.4578843261293\\
1.84928909550999	3.3236659478454\\
0.482414938556876	3.37466623621593\\
0.525686159671997	3.53244245123575\\
0.508295744358599	3.42963809924912\\
2.37812734702552	3.34029062505275\\
0.504982518714676	3.40799206139799\\
2.74539444135438	3.51479954590655\\
0.539439348300302	3.3128326985921\\
0.503013706173872	3.55623149684968\\
2.29122862124764	3.49830911367035\\
3.01110043976927	3.45928679089726\\
0.559748537610821	3.28986284180812\\
0.499745265277187	3.55051142958167\\
3.31923603011056	3.56287482505913\\
0.610177065624624	3.12976764924278\\
2.82591633868221	3.52071395137149\\
2.53844272285642	3.32281938519661\\
2.53185229537307	3.30045544206538\\
1.95393534094569	3.28563868381432\\
2.91019465115303	3.31610155629465\\
3.63922701165574	3.29370164480628\\
3.07792879365427	3.07682179931178\\
0.505026189640759	3.1336938370665\\
3.66899679108451	3.46691325348133\\
3.24848888100858	2.87808398259424\\
3.03719931805167	2.96804858969895\\
2.63527798392346	3.10525800482249\\
};
\addlegendentry{$\text{Intersection of }\phi{}_\text{T,P3}\text{ and }\phi{}_\text{R,P0}$}

\addplot[area legend, draw=black, fill=blue, fill opacity=0.3, forget plot]
table[row sep=crcr] {%
x	y\\
0	5.285\\
0	3.85\\
0.1	3.85\\
0.1	5.285\\
}--cycle;

\addplot[area legend, draw=black, fill=blue, fill opacity=0.3, forget plot]
table[row sep=crcr] {%
x	y\\
0	2.35\\
0	0.915\\
0.1	0.915\\
0.1	2.35\\
}--cycle;

\addplot[area legend, draw=black, fill=green, fill opacity=0.3, forget plot]
table[row sep=crcr] {%
x	y\\
5.07	6.4\\
5.07	1\\
6.03	1\\
6.03	6.4\\
}--cycle;
\node[right, align=left, inner sep=0]
at (axis cs:0.561,6.031) {P3\\P4};
\node[right, align=left, inner sep=0]
at (axis cs:4.693,5.916) {P0\\P1};
\node[right, align=left, inner sep=0]
at (axis cs:0.556,0.461) {P2\\P5};
\node[right, align=left, inner sep=0]
at (axis cs:4.589,0.431) {P6\\P7};
\node[right, align=left, inner sep=0, rotate=90]
at (axis cs:0.3,4.05) {Window};
\node[right, align=left, inner sep=0, rotate=90]
at (axis cs:0.3,1.115) {Window};
\node[right, align=left, inner sep=0, rotate=90]
at (axis cs:0.3,1.115) {Window};
\node[right, align=left, inner sep=0, rotate=90]
at (axis cs:5.55,1.7) {Server racks + acoustic panels};
\addplot [color=red, line width=2.0pt]
  table[row sep=crcr]{%
0.5	3.4\\
3.5	3.4\\
};
\addlegendentry{Walking trajectory}
\addplot[only marks, mark=square*, mark size=2.5pt, draw=black]table[row sep=crcr]{%
x	y\\
0.5     3.4\\};
\addlegendentry{Start}

\addplot[only marks, mark=otimes, mark size=2.5pt, draw=black]table[row sep=crcr]{%
x	y\\
3.5  3.4\\};
\addlegendentry{Stop}

\end{axis}
\end{tikzpicture}%